\documentclass[twocolumn,trackchanges]{aastex701}
\usepackage{amsmath}
\usepackage{amssymb}
\usepackage{bm}
\usepackage{graphicx}
\usepackage{multirow}
\usepackage{booktabs}

\newcommand{\model}{MAGIKS}
\newcommand{\kn}{kilonova}
\newcommand{\kns}{kilonovae}
\newcommand{\gw}{gravitational-wave}

\newcommand{\bns}{binary neutron star}
\newcommand{\nsbh}{neutron star-black hole}

\shorttitle{MAGIKS}
\shortauthors{B. Gao et al.}
\submitjournal{ApJS}
\begin{document}

\title{MAGIKS: Multimodal Alignment and Gravitational-Wave-Conditioned Interaction for Kilonova Search}

%% Author information. AASTeX v7+ requires an email address for every author;
%% use \email[show]{...} only for the corresponding author or other emails
%% that should be displayed on the title page.
\correspondingauthor{Jingqi Zhang, Linqing Wen, Wen Zhao}

\author[orcid=0009-0002-0900-9901,gname=Bingzhou,sname=Gao]{Bingzhou Gao}
\affiliation{Department of Astronomy, University of Science and Technology of China, Hefei, Anhui 230026, People's Republic of China}
\email{bingzhougao.mail@gmail.com}

\author[orcid=0000-0002-4320-4420,gname=Weichangfeng,sname=Guo]{Weichangfeng Guo}
\affiliation{Department of Physics, University of Western Australia, Perth, WA 6009, Australia}
\email{weichangfeng.guo@research.uwa.edu.au}

\author[orcid=0009-0009-7645-4299,gname=Jingqi,sname=Zhang]{Jingqi Zhang}
\affiliation{Department of Astronomy, University of Science and Technology of China, Hefei, Anhui 230026, People's Republic of China}
\affiliation{School of Astronomy and Space Sciences, University of Science and Technology of China, Hefei, Anhui 230026, People's Republic of China}
\email[show]{zhangjingqi@mail.ustc.edu.cn}

\author[orcid=0000-0001-5424-8368,gname=Alistair,sname=McLeod]{Alistair McLeod}
\affiliation{Department of Physics, University of Western Australia, Perth, WA 6009, Australia}
\email{alistair.mcleod@research.uwa.edu.au}

\author[orcid=0000-0002-2242-1514,gname=Zhengyan,sname=Liu]{Zhengyan Liu}
\affiliation{Department of Astronomy, University of Science and Technology of China, Hefei, Anhui 230026, People's Republic of China}
\affiliation{School of Astronomy and Space Sciences, University of Science and Technology of China, Hefei, Anhui 230026, People's Republic of China}
\email{ustclzy@mail.ustc.edu.cn}

\author[orcid=0000-0001-7987-295X,gname=Linqing,sname=Wen]{Linqing Wen}
\affiliation{Department of Physics, University of Western Australia, Perth, WA 6009, Australia}
\email[show]{linqing.wen@uwa.edu.au}

\author[orcid=0000-0002-1330-2329,gname=Wen,sname=Zhao]{Wen Zhao}
\affiliation{Department of Astronomy, University of Science and Technology of China, Hefei, Anhui 230026, People's Republic of China}
\affiliation{School of Astronomy and Space Sciences, University of Science and Technology of China, Hefei, Anhui 230026, People's Republic of China}
\affiliation{College of Physics, Guizhou University, Guiyang 550025, People's Republic of China}
\email[show]{wzhao7@ustc.edu.cn}

%% Use the \collaboration command to identify collaborations. This command
%% takes an optional argument that is either a number or the word "all"
%% which tells the compiler how many of the authors above the command to
%% show. For example "\collaboration[all]{(DELVE Collaboration)}" wil include
%% all the authors above this command.
%%
%% Mark off the abstract in the ``abstract'' environment.
\begin{abstract}

We present MAGIKS, a multimodal framework for gravitational-wave-triggered kilonova identification that formulates a counterpart search as query-conditioned gallery retrieval. MAGIKS jointly encodes gravitational-wave source parameters and sky-localization maps with irregularly sampled multiband optical light curves and sky coordinates, following an align-before-fuse design. We apply a time-aware supervised contrastive objective to provide coarse cross-modal alignment, followed by a dual cross-attention fusion module that ranks candidates by evaluating (1) the consistency between optical coordinates and the gravitational-wave sky-localization map and (2) the consistency between gravitational-wave source parameters and the optical temporal evolution.
We train and evaluate MAGIKS on simulated binary neutron star (BNS) and neutron star-black hole (NSBH) mergers, associated Rubin Observatory kilonova light curves generated under regular survey observations and target-of-opportunity follow-up, GW-detected mergers without a Rubin-detectable kilonova counterpart, and realistic non-kilonova contaminants. Using independent test galleries containing a single true counterpart among time-compatible and sky-compatible distractors, we show that incorporating gravitational-wave information substantially improves retrieval performance over optical-only and sky-localization-only searches, with the advantage increasing as the candidate pool grows. In the largest evaluated galleries, each containing 5000 candidates, MAGIKS ranks the true counterpart first for 39.8\% of queries, compared with 4.8\% for optical-only ranking and 0.2\% for sky-localization-only ranking. 
In an event-specific GW170817 study, we generate counterfactual Rubin observations using the released source parameters and three-dimensional localization, the measured counterpart position, and host distance. With 5000 candidates, MAGIKS ranks the true counterpart first for 87.3\% of queries and within the top ten for 95.7\%. In the pairwise-classification experiment, MAGIKS correctly classifies 94.2\% of the associated and non-associated GW-optical pairs. In a separate crossed-pair sensitivity experiment, its matching score responds significantly to controlled changes in all five tested GW parameters. These results support gravitational-wave-conditioned kilonova candidate prioritisation for future multi-messenger follow-up.

\end{abstract}

%% Keywords should appear after the \end{abstract} command.
%% The AAS Journals now uses Unified Astronomy Thesaurus (UAT) concepts:
%% https://astrothesaurus.org
%% You will be asked to selected these concepts during the submission process
%% but this old "keyword" functionality is maintained in case authors want
%% to include these concepts in their preprints.
%%
%% You can use the \uat command to link your UAT concepts back its source.
\keywords{
\uat{Gravitational waves}{678} ---
\uat{Transient sources}{1851} ---
\uat{Time domain astronomy}{2109} ---
\uat{Astronomy data analysis}{1858} ---
\uat{Neural networks}{1933}
}
% \begin{keywords}
% gravitational waves --- transients: neutron star mergers --- methods: data analysis --- methods: statistical --- techniques: photometric --- surveys
% \end{keywords}

%% From the front matter, we move on to the body of the paper.
%% Sections are demarcated by \section and \subsection, respectively.
%% Observe the use of the LaTeX \label
%% command after the \subsection to give a symbolic KEY to the
%% subsection for cross-referencing in a \ref command.
%% You can use LaTeX's \ref and \label commands to keep track of
%% cross-references to sections, equations, tables, and figures.
%% That way, if you change the order of any elements, LaTeX will
%% automatically renumber them.

\section{Introduction}

The detection of GW170817 and its optical counterpart established compact-binary mergers as a central target of multimessenger astronomy \citep{collaborationGW1708172017,abbottMultimessenger2017,coulterSwope2017}. The event demonstrated that \kns\ can connect \gw\ measurements to r-process nucleosynthesis, merger ejecta physics, and the electromagnetic evolution of neutron-star mergers \citep{metzgerKilonovae2019,radiceBinary2018}. It also exposed a practical limitation of current follow-up strategies: even for a nearby event, the counterpart must be found among tens to hundreds of optical transients distributed over an uncertain \gw\ localization region \citep{liuIlluminating2026,huKilonova2025,ahumadaSearching2024}.

This limitation will become more severe in the Rubin Observatory Legacy Survey of Space and Time (LSST) \citep{ivezicLSST2019,hambletonRubin2023} era. Rubin will produce a dense, high-volume stream of optical alerts, while future \gw\ observing runs are expected to increase the number of compact-binary triggers requiring electromagnetic follow-up due to detector upgrades. \citep{hambletonRubin2023,petrovDatadriven2022,shahPredictions2024}. For each \gw\ event, especially with poor localization, wide-field searches may return many time-compatible and sky-compatible candidates, including supernovae, active galactic nuclei, tidal disruption events, image-subtraction artifacts, and unrelated fast transients \citep{andreoniFasttransient2021,andreoniTargetofopportunity2022,stevensonStrategy2026}. Because spectroscopy and deep follow-up remain limited resources, counterpart identification is not only a detection problem but also a prioritization problem: candidates must be ranked rapidly and robustly for a specific \gw\ query.

Existing counterpart-search pipelines use combinations of \gw\ sky maps, galaxy catalogs, observing constraints, and transient-filtering heuristics to reduce the search space \citep{singerRapid2016,ducoinOptimizing2020}. Machine-learning methods have also been developed for kilonova filtering and target selection in alert streams \citep{stachieUsing2020,biswasEnabling2023,liangKilonovaTargeting2023}. These approaches either treat the optical candidate independently of the triggering \gw\ event or use the \gw\ information primarily as an external spatial prior. For example, the Fink kilonova science module uses a random-forest classifier based on optical light-curve features to identify kilonova candidates without conditioning its classification score on a specific triggering \gw\ event \citep{biswasEnabling2023}.

The difficulty of directly matching gravitational wave event data and optical data is intrinsically multimodal. Optical candidates are represented by irregular, heterogeneous, and often sparse multiband light curves, for which sequence models and early-time classifiers have proven useful in large transient-classification settings \citep{kesslerModels2019,muthukrishnaRAPID2019,2020MNRAS.491.4277M,biswasEnabling2023}. By contrast, a \gw\ trigger is represented by uncertain source parameters and a probabilistic sky-localization map of typically tens to hundreds of square degrees. Optical evolution can indicate whether a transient resembles a \kn, but it cannot by itself test whether the transient is spatially and physically compatible with a particular \gw\ event. Conversely, sky localization coincidence alone provides useful filtering, but within the typical localization of a gravitational wave event, there could be tens to hundreds of optical transients or host galaxy candidates.

Recent astronomical multimodal models have begun to demonstrate how heterogeneous observations can be mapped into shared representation spaces. AstroCLIP uses contrastive learning to align imaging and spectroscopic information, enabling cross-modal retrieval and physically informative embeddings for galaxies \citep{parkerAstroCLIP2024}. AstroM$^3$ extends this direction by combining photometry, spectra, and metadata in a unified astronomical representation-learning framework \citep{rizhkoAstroM32025}, while MAVEN explores multimodal learning for transient and variable-source characterisation by combining light-curve information with contextual observables \citep{zhangMaven2024}. These studies show that multimodal pre-training can extract complementary information from different astronomical data products. The \gw-optical counterpart-search problem is nevertheless distinct: it requires matching a specific uncertain \gw\ trigger to a ranked gallery of optical candidates, where spatial consistency, transient evolution, and physical compatibility must be evaluated jointly.

In this paper, we formulate \gw-triggered \kn\ identification as a gallery retrieval task for Rubin baseline and target-of-opportunity (ToO) observations. The simulated observations span baseline, Gold, and Silver classes selected according to the recovered network signal-to-noise ratio (SNR) and the 90\% credible localization area. The baseline class uses Wide-Fast-Deep (WFD) and Deep Drilling Fields (DDF) observations, and the Gold and Silver classes combine these observations with simulated ToO follow-up \citep{andreoniRubin2024}. Given one counterpart-bearing \gw\ event as the query and a pool of optical transients as the gallery, the goal is to rank the true counterpart as high as possible. An auxiliary pairwise-classification task also includes \gw-detected mergers without a Rubin-detectable \kn\ counterpart. We introduce \model, a multimodal architecture following the align-before-fuse strategy \citep{liAlign2021} and organized around the physical structure of \gw-optical association. The model combines a dual-stream \gw\ encoder for scalar descriptors and sky-localization maps with a transformer-based optical encoder for multiband light curves. A time-aware supervised contrastive objective first aligns the two modalities, and a dual cross-attention fusion module then computes pairwise matching scores for candidate ranking by allowing optical coordinates to query the \gw\ skymap and \gw\ descriptors to query optical light curves.

\begin{figure*}[!tp]
    \centering
    \includegraphics[width=\linewidth]{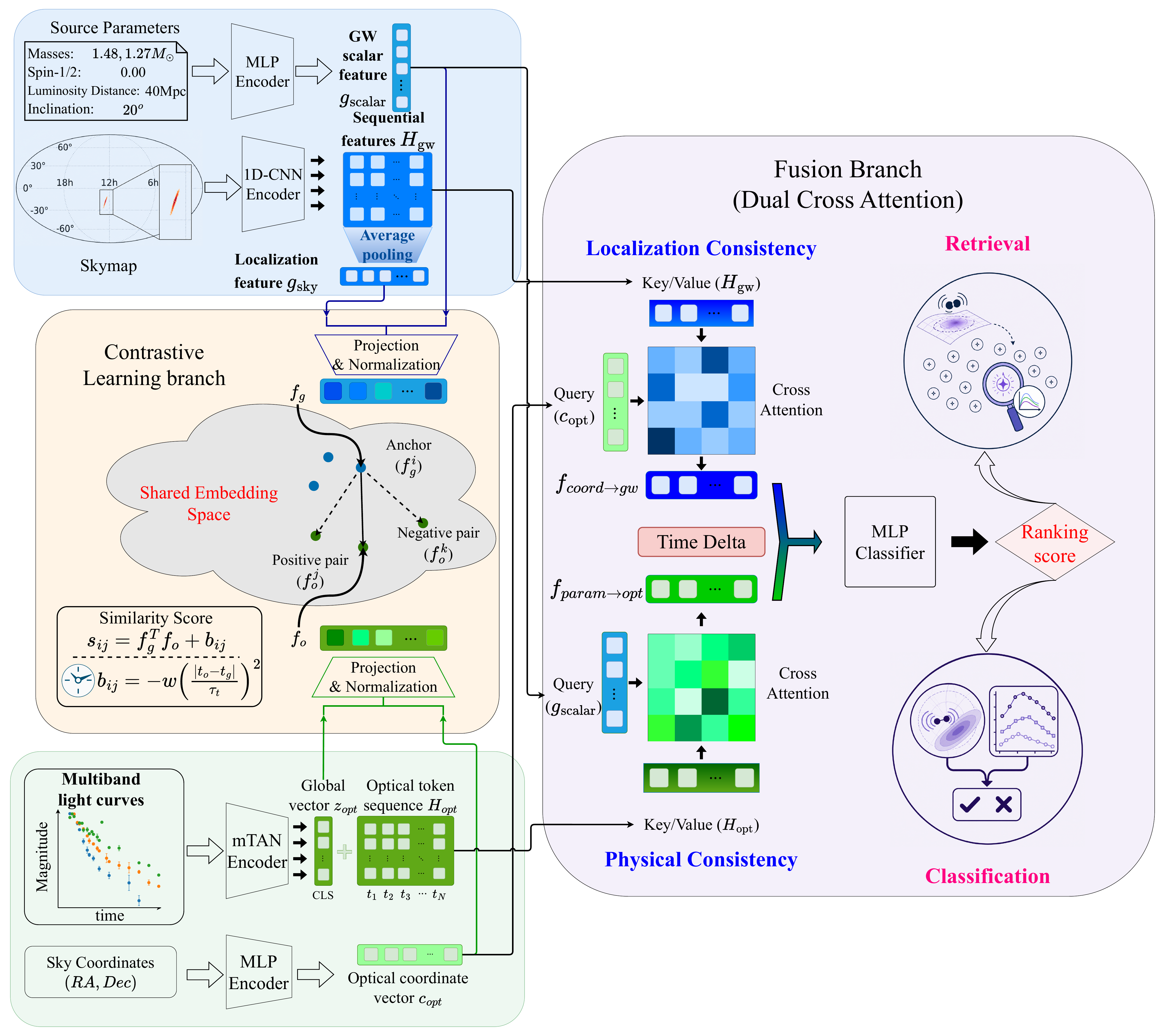}
    \caption{Overview of the proposed multimodal framework for gravitational-wave-triggered kilonova counterpart search. The upper stream encodes the gravitational-wave trigger using source-parameter descriptors and the sky-localization map: scalar parameters are projected with a multilayer perceptron (MLP), while the skymap is processed by a one-dimensional convolutional network and pooled into a localization feature. The lower stream encodes Rubin/LSST optical alerts using irregularly sampled multiband light curves and sky coordinates: temporal samples are represented on a reference time grid with time embeddings and encoded by the Multi-Time Attention Network (mTAN) module, while coordinates are encoded separately. The contrastive-learning branch projects the gravitational-wave and optical representations into a shared embedding space and uses a supervised contrastive objective with a time-compatibility term to favour physically plausible pairs. The fusion branch performs two explicit cross-modal matching operations, comparing optical coordinates with the gravitational-wave localization and gravitational-wave source descriptors with the optical temporal evolution, and produces the final ranking score for counterpart retrieval.}
    \label{fig:method_overview}
\end{figure*}

This work addresses a specific question: given a gravitational-wave trigger and a set of optical transients that have passed temporal and sky-consistency cuts, can a learned multimodal score improve the ranking of the true kilonova counterpart? Our approach has three components. First, we cast gravitational-wave-triggered counterpart search as a query-conditioned retrieval problem. Second, we retain two physically distinct matching operations: optical coordinates are compared with the gravitational-wave skymap, and gravitational-wave source descriptors are compared with the optical temporal evolution. Third, we evaluate the framework through retrieval and classification ablations, an event-specific GW170817 experiment using the released source parameters, localization, and distance information, and a controlled analysis of sensitivity to individual gravitational-wave parameters. Across these evaluations, incorporating gravitational-wave information provides a clear advantage in large candidate galleries. In the independent simulated test galleries containing 1000 candidates, \model\ ranks the true counterpart first for 54.3\% of queries, compared with 21.1\% for optical-only ranking and 0.7\% for skymap-only ranking. For GW170817 at the same gallery size, the corresponding fraction reaches 93.1\%, while in the separate pairwise-classification experiment \model\ correctly classifies 94.2\% of associated and non-associated pairs. Together, these results demonstrate the value of jointly modeling gravitational-wave and optical information for prioritizing counterparts in crowded candidate galleries. A schematic overview of \model\ is shown in Figure~\ref{fig:method_overview}.
% The present study is therefore intended as a controlled simulation benchmark and does not constitute a complete low-latency broker pipeline.

\clearpage

\section{Simulated Multimodal Dataset}

\subsection{Gravitational-wave event simulation}
\label{sec:gw_sim}

The \gw\ samples are generated with the GWSamplegen package \citep{mcleodBinary2025}\footnote{\url{https://github.com/alistair-mcleod/GWSamplegen}}. Compact binary coalescence signals are injected into Gaussian noise realizations for the Hanford-Livingston-Virgo (HLV) detector network under an O5a sensitivity model \citep{fritschelO5Strain2025,rocchiPublicO5Virgo2025}. All three HLV detectors are treated as operating in the sample generation process. For each simulated source, GWSamplegen performs a simplified matched-filter analysis over a local subset of an aligned-spin template bank, generating detector SNR time series from which simulated event triggers are constructed. Rejection sampling is used to ensure that only events with a recovered network SNR of at least 8 enter the production samples. For each accepted event, the template yielding the highest recovered network SNR supplies the recovered component masses and aligned spins. These recovered parameters and the detector SNR time series are passed to BAYESTAR \citep{singerRapid2016} to construct a three-dimensional multi-order HEALPix localization map \citep{gorskiHEALPix2005}, which is sampled into a fixed pixel sequence.

The \gw\ information used by \model\ is separated into two categories. (1) The three-dimensional sky-localization map recovered with BAYESTAR provides the spatial information associated with each \gw\ event. (2) The seven scalar inputs comprise the detector-frame component masses and aligned-spin components recovered from the best-matching template, the injection-derived $\cos\iota$, and the mean and standard deviation of the BAYESTAR luminosity-distance posterior. For application to observed events, we assume access to source-parameter inference products in addition to the publicly distributed sky-localization and distance information. These include component-mass estimates from the best-matching search template and posterior samples for the component masses, inclination, sky position, and luminosity distance. Such products can be obtained through collaboration-internal LVK analyses, including online Bilby parameter estimation, once the relevant analyses are complete \citep{ashtonBilby2019,morisakiRapid2023}, or from published GW catalogs and their accompanying posterior samples in offline studies \citep{ligoscientificcollaborationandvirgocollaborationGWTC12019}. In the present simulations, using the injected inclination provides an idealized reference with perfectly known inclination. We examine the effect of inclination errors on retrieval performance in Section~\ref{sec:discussion}.

For training and validation, component masses and aligned spins are drawn from uniform proposal distributions, while binary orientations are isotropic. The retained distributions are shaped by the ejecta and GW-selection criteria. A different sampling strategy is adopted for the test set. For \bns\ mergers, the component masses are drawn uniformly from $[1.0, 2.05]\,M_{\odot}$ and the dimensionless aligned spins from $[-0.1, 0.1]$. We adopt mass limits of neutron stars consistent with the equation of state (EOS) used in the ejecta prescription. Specifically, we use ejecta-fitting relations calibrated for the SFHo EOS \citep{steinerCORECOLLAPSE2013} for the light curve simulation \citep{setzerModelling2023,shahPredictions2024}. For \nsbh\ mergers, neutron-star parameters are sampled from the same ranges, while black-hole masses and aligned spins are drawn from $[2.05, 10]\, M_{\odot}$ and $[-0.99, 0.99]$, respectively. After drawing candidate source parameters, we calculate the ejecta properties for both BNS and NSBH systems using the corresponding fitting formulae described in Section~\ref{ch:optical_data}. We then apply rejection sampling based on these calculated ejecta properties, accepting only proposals for which both the dynamical and wind ejecta masses are strictly positive and lie within their respective ranges supported by the optical model \citep{bullaPOSSIS2019,kesslerSNANA2009}. The accepted source-parameter samples enter the GW simulation catalogs and are subsequently used in the matched-filter recovery and BAYESTAR localization stages described above. Distances are sampled uniformly in volume up to 500~Mpc for BNS and 1000~Mpc for NSBH. In addition, we assumed an isotropic distribution of sources and binary orientations. Sky positions were generated with uniform right ascension and uniform sin(declination), while the inclination angle $\iota$ was generated from a uniform distribution in $\cos\iota$ over $[-1, 1]$.

\begin{figure}
    \centering
    \includegraphics[width=1\linewidth]{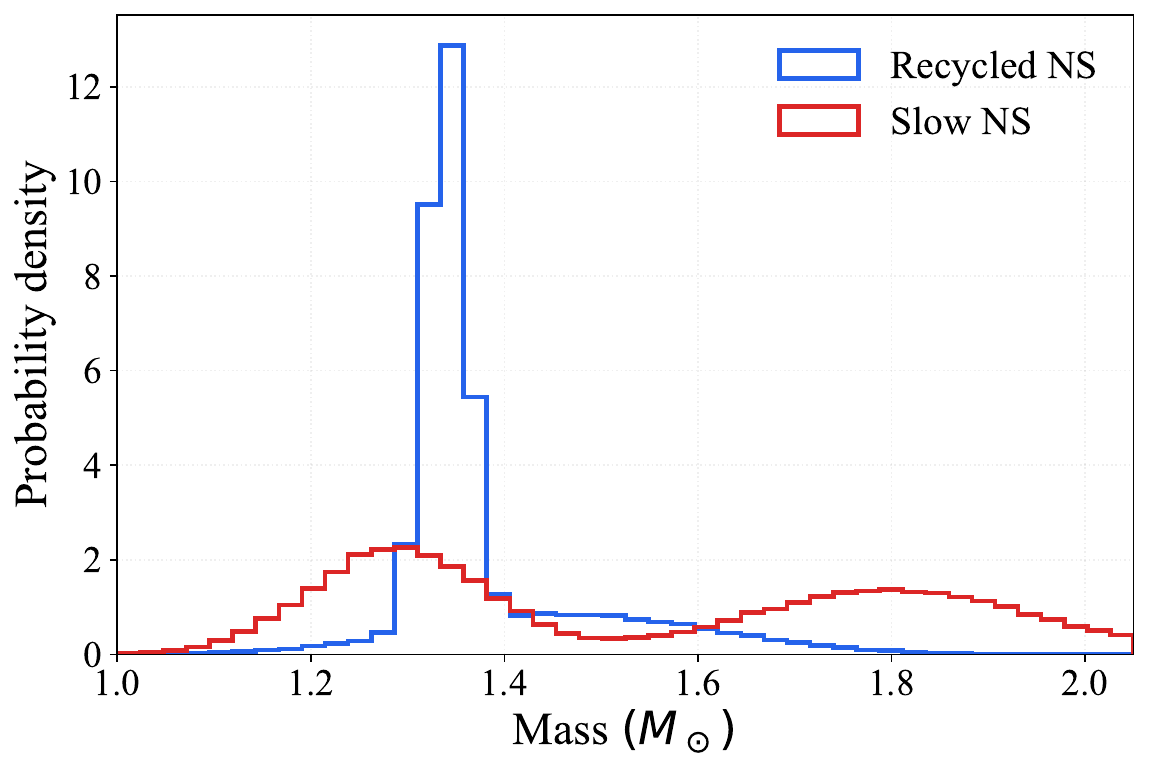}
    \caption{Sampled component-mass distributions for the BNS test set. The double-Gaussian distribution of recycled neutron stars (blue) is from \citet{farrowMass2019}, and the distribution of slowly rotating neutron stars (red) is from \citet{galaudageHeavy2021}. Both components are truncated to the neutron-star mass range used in this work.}
    \label{fig:bns_param_test}
\end{figure}

\begin{figure}
    \centering
    \includegraphics[width=1\linewidth]{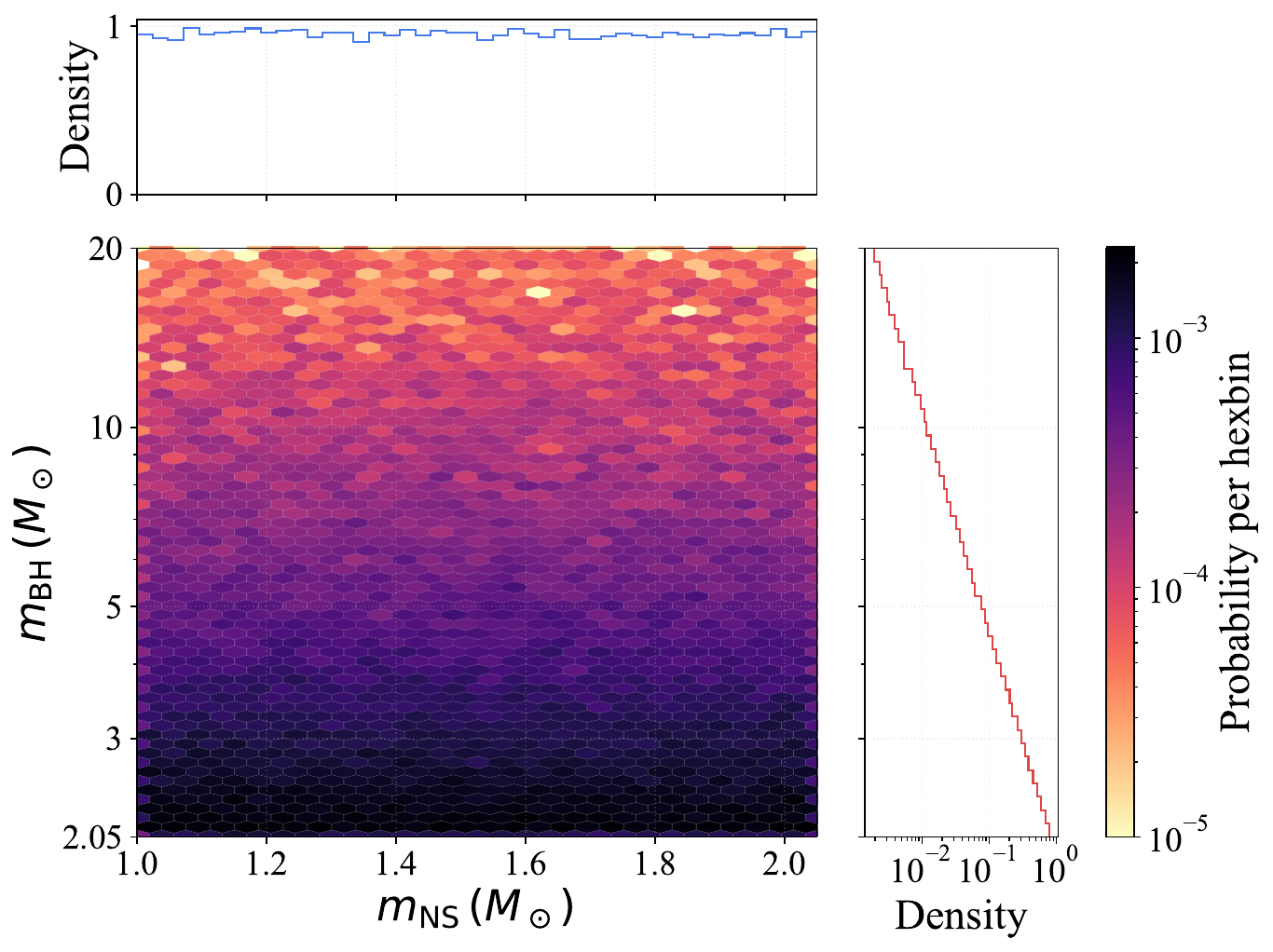}
    \caption{Joint distribution of sampled component masses for the NSBH test set. The central panel shows the two-dimensional distribution of neutron-star mass $m_{\rm NS}$ and black-hole mass $m_{\rm BH}$, with colour indicating the probability per hexagonal bin. The top and right panels show the distributions of neutron stars (NS) and black holes (BH) separately. The BH mass axis is logarithmic.}
    \label{fig:nsbh_param_test}
\end{figure}

For independent testing, we adopt astrophysically motivated component-mass distributions. For BNS systems, we assume the standard formation channel, where each binary consists of a first-born recycled neutron star with mass $M_{\mathrm{recycled}}$, spun up by accretion, and a second-born slowly rotating neutron star with mass $M_{\mathrm{slow}}$. Based on prior research \citep{galaudageHeavy2021,farrowMass2019}, we used a double-Gaussian distribution to model the masses of each component, as summarized in Figure~\ref{fig:bns_param_test}. For NSBH systems, we adopt uniform and power-law distributions for the masses of neutron stars and black holes separately \citep{zhuPopulation2022}, as shown in Figure~\ref{fig:nsbh_param_test}. The aligned spin and extrinsic parameters are sampled using the same distributions as in the training--validation set, except for distance which is limited to 500~Mpc. Table~\ref {tab:astro_test_distribution} summarizes the astrophysical distributions for the test dataset, and the distribution parameters follow the studies cited above.
% Figure~\ref{fig:bns_param_test} shows the component mass distribution for \bns , and Figure~\ref{fig:nsbh_param_test} shows the joint distribution of masses for \nsbh .

Following the ejecta selection and the requirement of a recovered network SNR of at least 8 during GW recovery, the GW simulation catalogs contain 20,000 BNS and 20,000 NSBH events for training and validation, and 2,000 BNS and 2,000 NSBH events for independent testing. All retained events also have valid BAYESTAR localization products. Separate type-1 ejecta-free catalogs contain 5,000 BNS and 5,000 NSBH events for training and validation, and 500 BNS and 500 NSBH events for testing. Type-2 optically undetected samples are identified from the GW simulation catalogs after the Rubin simulations in Section~\ref{ch:optical_data}. The final positive and negative GW sample counts are given in Section~\ref{ch:dataset_split}.

\begin{deluxetable*}{@{}c@{\hspace{1.6em}}c@{\hspace{1.6em}}c@{\hspace{1.9em}}c@{\hspace{1.9em}}c@{}}
\tablewidth{\textwidth}
\tablecaption{Mass distributions adopted for the astrophysical test set \label{tab:astro_test_distribution}}
\tablehead{
\colhead{Type} & \colhead{Component} & \colhead{Distribution} & \colhead{Parameters} & \colhead{Mass range}
}
\startdata
\multirow{4}{*}{BNS} & \multirow{2}{*}{Recycled NS} & \multirow{2}{*}{Double Gaussian} & $w_1=0.68,\ \mu_1=1.34,\ \sigma_1=0.02$ & \multirow{2}{*}{$[1.0,2.05]\,M_{\odot}$} \\
 &  &  & $w_2=0.32,\ \mu_2=1.47,\ \sigma_2=0.15$ &  \\
 & \multirow{2}{*}{Slow NS} & \multirow{2}{*}{Double Gaussian} & $w_1=0.50,\ \mu_1=1.29,\ \sigma_1=0.09$ & \multirow{2}{*}{$[1.0,2.05]\,M_{\odot}$} \\
 &  &  & $w_2=0.50,\ \mu_2=1.80,\ \sigma_2=0.15$ &  \\
\hline
\multirow{2}{*}{NSBH} & BH & Power law & $\alpha=2.7$ & $[2.05,20]\,M_{\odot}$ \\
 & NS & Uniform & \nodata & $[1.0,2.05]\,M_{\odot}$ \\
\enddata
\tablecomments{The columns list the compact-binary type, the component to which the distribution is applied, the functional form, the distribution parameters, and the allowed mass range. For the BNS double-Gaussian distributions, $w$, $\mu$, and $\sigma$ denote the weight, mean, and standard deviation of each Gaussian component. For NSBH systems, black-hole masses follow $P(m_{\rm BH})\propto m_{\rm BH}^{-\alpha}$.}
\end{deluxetable*}

\subsection{Optical transient simulation}
\label{ch:optical_data}

Optical samples are simulated \kn\ light curves generated for the \gw\ events described in Section~\ref{sec:gw_sim}. For each training--validation event, we randomly draw $1000$ joint sky-position and luminosity-distance samples from the three-dimensional posterior probability density of its BAYESTAR map, with the sky positions restricted to the 90\% credible region. This fixed-size posterior sampling exposes the light-curve simulation to different Rubin cadences and observing conditions. For each test event, we draw 63 samples from the same three-dimensional posterior and additionally include the injected sky position and distance as the physically consistent counterpart, giving at most 64 light curves per event.

We further derive ejecta properties relevant to kilonova light curves from the injected binary parameters using astrophysical fitting formulae for BNS \citep{shahPredictions2024,setzerModelling2023,coughlinMultimessenger2019} and NSBH \citep{gompertzMultimessenger2023a,foucartRemnant2018,krugerEstimates2020,raaijmakersChallenges2021}, and simulate Rubin-like observations with SNANA \citep{kesslerSNANA2009} using the POSSIS \kn\ model \citep{bullaPOSSIS2019}. The observing plans combine the WFD and DDF observations from the Rubin baseline v5.1 cadence \citep{yoachimLsstsims2026} with simulated ToO follow-up. Following the Rubin ToO strategy \citep{andreoniRubin2024}, events are assigned to baseline, Gold, or Silver observing classes according to their network SNR and 90\% credible localization area, $A_{90}$. Events with a network SNR less than 12 or $A_{90}\geq500\,\mathrm{deg}^{2}$ use the baseline WFD and DDF cadence. Among events with a network SNR above 12, those with $A_{90}\leq100\,\mathrm{deg}^{2}$ receive Gold ToO observations, while those with $100<A_{90}<500\,\mathrm{deg}^{2}$ receive Silver ToO observations. In the training--validation dataset, around 45.9\% of the positive \gw\ events receive ToO follow-up, and these events contribute about 89.6\% of the optical light curves.

\begin{figure}
    \centering
    \includegraphics[width=\linewidth]{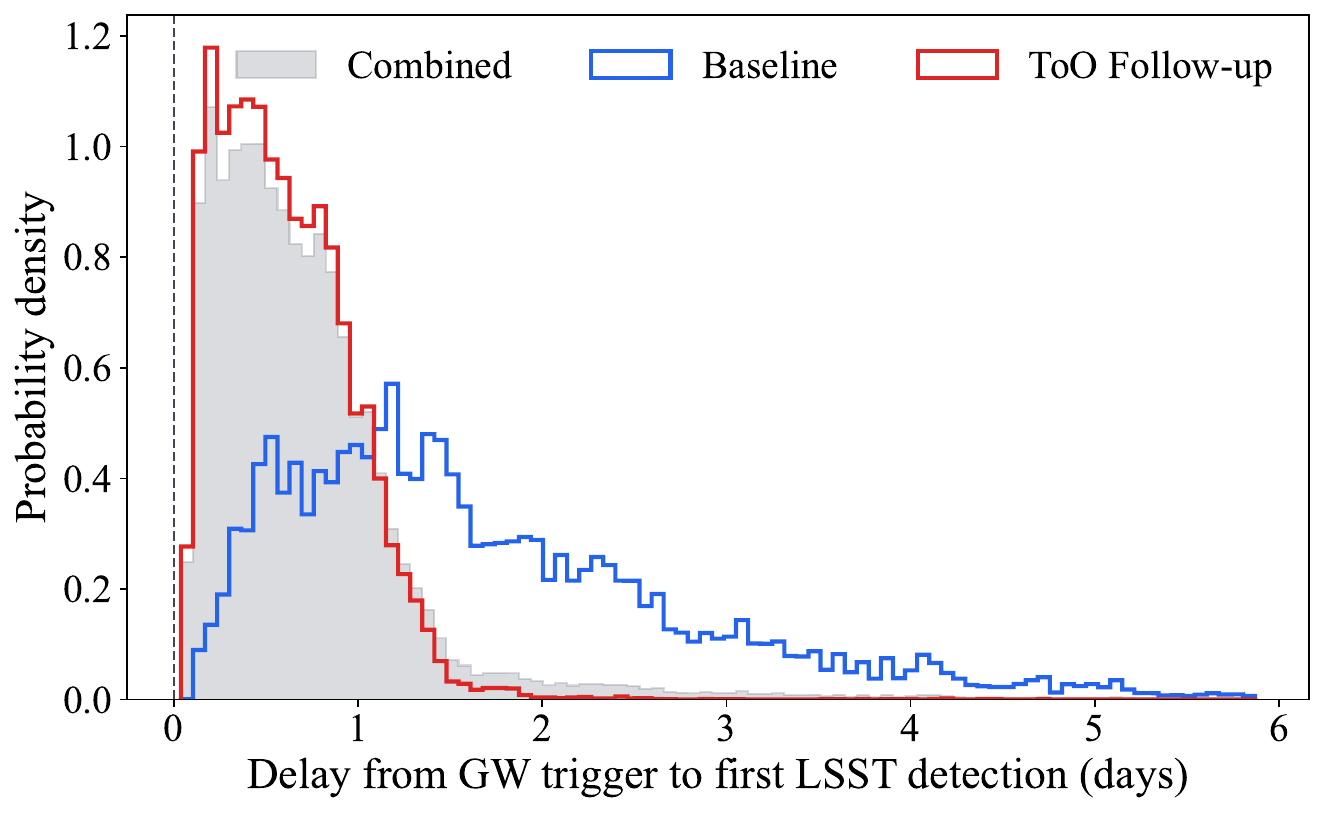}
    \caption{\added{Empirical probability-density distributions of the delay between the gravitational-wave trigger time and the first retained LSST detection in the training--validation dataset. The blue curve shows the baseline light curves, the red curve shows the light curves with ToO follow-up, and the shaded gray curve shows their combined distribution. ToO follow-up is applied to 45.9\% of the positive gravitational-wave events and contributes 89.6\% of the retained light curves. The median delays are 1.43 days for Baseline, 0.56 days for ToO follow-up, and 0.60 days for the combined sample.}}
    \label{fig:delay_time}
\end{figure}

To reduce the mismatch between simulated and observationally formatted light curves, we preprocess optical data in three steps. First, same-band measurements within a two-hour window are merged by inverse-variance weighting. In addition, to keep only practically useful data, we define optical detections as merged Rubin observations with SNR $\ge5$, the same criterion as LSST alerts \citep{ivezicLSST2019}, and we require each light curve to contain at least two optical detections and five merged measurements. However, we retain all photometric measurements after two-hour same-band coaddition in the model input, including low-SNR and negative-flux points. Second, fluxes are represented in asinh magnitudes, also known as luptitudes \citep{luptonModified1999}; this representation keeps low-S/N and negative-flux points from difference imaging numerically well behaved. Third, optical zero time is anchored to the first detection, with its delay from the merger determined by the observing cadence. The delay distributions for the Baseline and ToO follow-up samples in the training-validation dataset, together with their combined distribution, are shown in Figure~\ref{fig:delay_time}. The median delay decreases from 1.43 days for the Baseline sample to 0.56 days for the ToO follow-up sample, while the combined sample has a median delay of 0.60 days.

The luptitude representation is
\begin{equation}
\begin{aligned}
\mu_{\rm lupt} &=
31.4-\frac{2.5}{\ln 10}\left[\operatorname{asinh}\left(\frac{f}{2b}\right)+\ln b\right],\\
\sigma_{\mu} &\approx
\frac{2.5}{\ln 10}\frac{\sigma_f}{\sqrt{f^2+(2b)^2}}.
\end{aligned}
\label{eq:luptitude}
\end{equation}
Here, $\mu_{\rm lupt}$ and $\sigma_\mu$ denote the luptitude and its uncertainty, and $f$, $\sigma_f$, and $b$ are expressed in nJy. We adopt $b=f_{5\sigma}/5$, with fixed band-dependent limiting depths; 31.4 is the corresponding AB zero point for flux densities in nJy.

\begin{figure*}
    \centering
    \includegraphics[width=\linewidth]{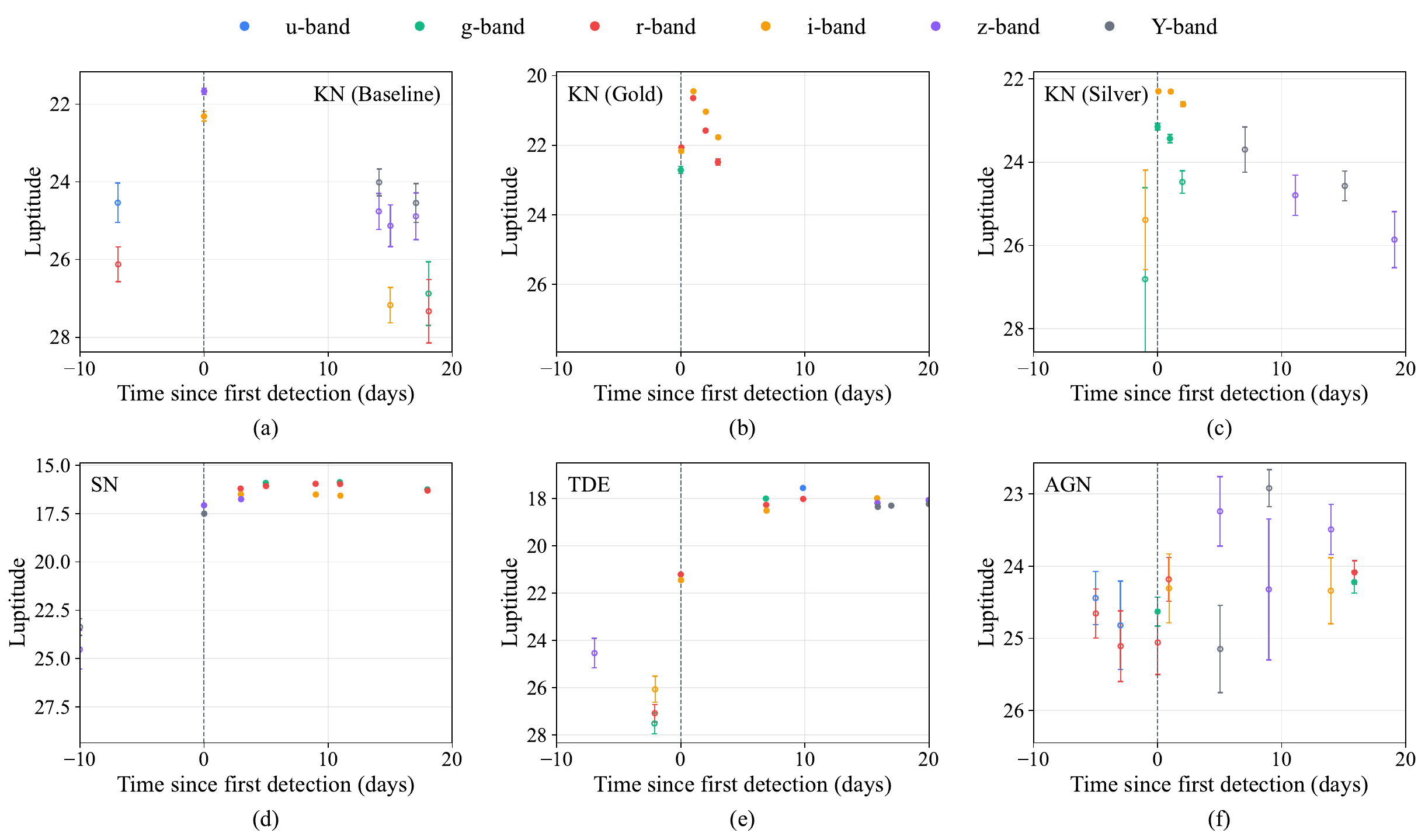}
    \caption{Representative \kn\ and non-\kn\ light curves in luptitude space after Rubin-style sampling and preprocessing. \added{Panels (a)--(c) show typical \kn\ light curves under the Baseline, Gold, and Silver observation strategies, respectively, while panels (d)--(f) show representative supernova, tidal disruption event, and active galactic nucleus contaminants, respectively.} Different colours denote the LSST $ugrizY$ bands, and the horizontal axis is measured relative to the first optical detection. Open circles represent non-detected observations, while filled circles represent detected observations. A detection is defined as having a signal-to-noise ratio (SNR) of at least 5.}
    \label{fig:lightcurves}
\end{figure*}

Non-\kn\ contaminants are drawn from Rubin-like optical transient simulations. ELAsTiCC2 \citep{elasticc_web} provides the optical contaminants for training and validation, while the independent test set uses ELAsTiCC \citep{narayan2023elasticc} contaminants. They provide negative samples and gallery distractors in the classification and retrieval tasks. Typical optical light curves with LSST sampling cadence are shown in Figure~\ref{fig:lightcurves}. The first row compares representative \kn\ light curves under the Baseline, Gold, and Silver observation strategies, while the second row shows representative SN, TDE, and AGN contaminants.

\subsection{Dataset splits}
\label{ch:dataset_split}

After generating the GW and optical data, we construct the training--validation and test datasets as follows. For each GW event with a detectable counterpart, we retain at most 1000 associated light curves in the training--validation dataset and at most 64 in the test dataset. Within this construction, GW events with at least one retained Rubin-detectable \kn\ light curve are treated as positive GW samples. We also include two classes of negative GW samples, defined as GW-detected BNS or NSBH mergers that lack a Rubin-detectable \kn\ counterpart in the dataset. The type-1 ejecta-free class consists of GW-detected mergers with zero dynamical and wind ejecta and therefore no predicted \kn\ emission. The type-2 optically undetected class is identified from the GW simulation catalogs and consists of mergers with nonzero ejecta and successful Rubin/SNANA coverage, but without any usable light curve satisfying the optical-detection criteria described in Section~2.2.

The training--validation dataset is split into training and validation subsets in an 8:2 ratio at the level of independent GW event IDs, and the sample drawn from the astrophysical distributions is used as the independent test set. After the GW and optical selections, the training--validation dataset contains $21,890$ positive GW samples and $19,814$ negative GW samples. The independent test dataset contains $1,843$ positive GW samples and $2,158$ negative GW samples, giving $4,001$ GW events in total. The resulting dataset sizes are summarized in Table~\ref{tab:dataset}.

\begin{deluxetable*}{lcccccc}
\tablewidth{\textwidth}
% \tabletypesize{\small}
\tablecaption{Summary of the multimodal datasets used for training, validation, and testing. \label{tab:dataset}}
\tablehead{
\colhead{} & \multicolumn{3}{c}{Training--validation set} & \multicolumn{3}{c}{Test set} \\
\colhead{Data type} & \colhead{BNS} & \colhead{NSBH} & \colhead{ELAsTiCC2} & \colhead{BNS} & \colhead{NSBH} & \colhead{ELAsTiCC}
}
\startdata
Light curves & \multicolumn{2}{c}{9,057,461} & 1,298,291 & \multicolumn{2}{c}{64,724} & 1,485,907 \\
Positive GW samples & 11,666 & 10,224 & \nodata & 956 & 887 & \nodata \\
Type-1 negative GW samples (zero ejecta) & 5,000 & 5,000 & \nodata & 500 & 500 & \nodata \\
Type-2 negative GW samples (optically undetected) & 4,814 & 5,000 & \nodata & 658 & 500 & \nodata \\
\enddata
\tablecomments{The kilonova light-curve counts are combined across BNS and NSBH sources, while the GW-event counts are reported separately by source class. Positive GW samples have at least one usable Rubin-detectable kilonova light curve. Type-1 samples are drawn from separate zero-ejecta catalogs. Type-2 samples are identified from the GW simulation catalogs after optical simulation and form a subset of those cataloged events. The difference between the sum of positive and type-2 samples and the full GW simulation catalogs mainly arises from events without successful Rubin coverage and the imposed cap on the number of type-2 negative samples. ELAsTiCC2 and ELAsTiCC sources are included only as optical contaminants and have no corresponding GW events.}
\end{deluxetable*}

\section{Method}

\subsection{Problem formulation}

The main task is \gw-triggered counterpart retrieval. For each \gw\ query event, there is a gallery containing one matched optical counterpart and $N-1$ distractors sampled from optical candidates associated with other events and other types of transients. In addition to retrieval, there is also a classification task to evaluate pairwise matching performance and identify the sources of false positives.

Let $\mathcal{V}_{\rm gw}^{+}$ and $\mathcal{V}_{\rm gw}^{-}$ denote the positive and negative \gw\ event sets, respectively, and let $\mathcal{V}_{\rm opt}$ denote the optical candidate set. A retrieval query $g_i\in\mathcal{V}_{\rm gw}^{+}$ is characterized by scalar descriptors and a localization map, while an optical candidate $o_j\in\mathcal{V}_{\rm opt}$ is characterized by its sky coordinates and an irregularly sampled multiband light curve. The goal is to assign a score or ranking statistic $l(g_i,o_j)$ such that the true counterpart is ranked near the top of the gallery for each query. Our approach consists of a cross-modal alignment stage that places likely matches close in embedding space and a fusion stage that computes a query-conditioned pair score for more accurate reranking.

\subsection{Overview of \model}

As shown in Figure~\ref{fig:method_overview}, \model\ combines gravitational-wave and optical encoders with alignment and fusion branches. The \gw\ encoder (upper left block) produces a global event representation for alignment, a scalar query representation for fusion, and a sequence of sky-localization tokens. The optical encoder (bottom left block) produces a global light-curve representation, temporal tokens over a fixed reference grid, and a coordinate embedding. The encoder outputs feed the alignment and fusion branches. The alignment branch (left middle block) uses supervised contrastive learning to pull coincident \gw-optical pairs together, while the fusion branch (right block) uses dual cross-attention to compute query-conditioned match scores.

\subsection{Gravitational-wave encoder}

The \gw\ encoder has dual-stream branches. A scalar branch takes the seven inputs defined in Section~\ref{sec:gw_sim}: the detector-frame component masses and aligned-spin components recovered from the best-matching template, the injection-derived $\cos\iota$, and the mean and standard deviation of the BAYESTAR luminosity-distance posterior. A second branch encodes the sky map as a sequence with seven channels per pixel: unit-sphere coordinates, pixel area, probability, and parameters of the distance distribution. These two types of input are fed into a multilayer perceptron (MLP) encoder and a one-dimensional convolutional network (1D-CNN) encoder separately.

The encoder outputs three quantities: a scalar-only feature $\bm{g}_{\rm param}\in\mathbb{R}^{D}$ used as a query in fusion, a sequence $\bm{H}_{\rm gw}\in\mathbb{R}^{L_g\times D}$ that preserves localization structure for optical-coordinate querying, and a \gw\ skymap embedding $\bm{g}_{\rm sky}\in\mathbb{R}^{D}$ obtained by applying global average pooling to $\bm{H}_{\rm gw}$.

\subsection{Optical encoder}
\label{cp:op_enc}

Irregular and multiband optical sequences are fed into a Multi-Time Attention Network (mTAN; \citealt{shuklaMultiTime2020}) encoder augmented in three ways. First, time is embedded with learnable periodic functions initialized over physically relevant periods of 0.5--100 days \citep{metzgerKilonovae2019,bullaPOSSIS2019,setzerModelling2023}. Second, we prepend a learnable classification (CLS) token to the reference-time queries to obtain a global light-curve representation alongside the temporal sequence. Third, the attention score \citep{vaswaniAttention2017} between reference time $t$ and observation time $t_{ld}$ includes a data-quality bias derived from flux-space SNR:
\begin{equation}
\begin{aligned}
\tilde{s}_{h}(t,t_{ld})
&=
\frac{(\mathbf{W}\phi_h(t))^{\intercal}(\mathbf{V}\phi_h(t_{ld}))}{\sqrt{d_k}} \\
&\quad +
\lambda \log \left[1+\operatorname{Softplus}\left(\mathrm{SNR}_{ld}-s_0\right)\right],
\end{aligned} 
\label{eq:mtan_snr}
\end{equation}
where $\operatorname{Softplus}(x) = \log\left(1 + e^x\right)$, $\phi_h$ is the periodic time embedding, $h$ indexes the attention head, and $\mathrm{SNR}_{ld}$ is the per-band S/N clipped between -8 and 20 at observation time $t_{ld}$. The reference times $t$ are placed on a uniformly spaced grid spanning $[-10,20]$ days. $\mathbf{W}$ and $\mathbf{V}$ are learnable parameters for projection, $\lambda$ is a learnable weighting parameter, and $s_0$ is the typical threshold set as 3. The first term in Equation~\ref{eq:mtan_snr} is the basic attention score of mTAN encoder, and the second term serves as the SNR-based bias. This mechanism improves the contribution of observations with high quality, and downweights less reliable measurements without discarding them entirely.

The optical branch returns a global CLS embedding $\bm{z}_{\rm opt}\in\mathbb{R}^{H}$, temporal tokens $\bm{H}_{\rm opt}\in\mathbb{R}^{L_o\times H}$ sampled on a fixed reference grid, and a coordinate embedding $\bm{c}_{\rm opt}\in\mathbb{R}^{K}$ obtained from the transient sky position. The dimensions of the GW and optical representations and their token-sequence lengths used in the main experiments are summarized in Table~\ref{tab:param_value}. The GW scalar and skymap embeddings are 128-dimensional, while the optical CLS embedding and temporal tokens are 192-dimensional. The optical and GW encoders produce sequences of 64 and 300 tokens, respectively.

With the optical encoder and another MLP classification block, we built an optical classification model that served as an optical-only baseline method. The baseline shared the same optical encoder backbone so that the comparison isolated the marginal value of \gw\ conditioning.

\subsection{Time-aware contrastive alignment}

The alignment branch follows the align-before-fuse philosophy of ALBEF \citep{liAlign2021}, but the loss is adapted to the many-to-one structure of a \gw\ counterpart search. We first project the global \gw\ and optical features into a shared embedding space and normalize them:
\begin{align}
\bm{f}_{g,i} &=
\operatorname{Norm}\bigl(
\psi_g([\bm{g}_{{\rm param},i};\bm{g}_{{\rm sky},i}])
\bigr),\\
\bm{f}_{o,j} &=
\operatorname{Norm}\bigl(
\psi_o([\bm{z}_{{\rm opt},j};\bm{c}_{{\rm opt},j}])
\bigr).
\end{align}
where $[\bm{a};\bm{b}]$ and $\psi()$ represent concatenation and projection. $\bm{f}_{g,i}\in\mathbb{R}^{J}$ and $\bm{f}_{o,j}\in\mathbb{R}^{J}$ represent the \gw\ feature of event $i$ and the optical feature of optical sample $j$, which encode both modalities in a shared latent space for contrastive alignment. The contrastive projections of both modalities are 128-dimensional.

The similarity score is computed as the dot product\citep{radfordLearning2021} with an additional temporal-compatibility bias:
\begin{equation}
s_{ij}=
\frac{\bm{f}_{g,i}^{\intercal}\bm{f}_{o,j}}{T}
-w\left(\frac{|t_{o,j}-t_{g,i}|}{\tau_t}\right)^2,
\label{eq:time_bias}
\end{equation}
where $T$ is the learnable temperature parameter used to rescale the dot-product similarity, with smaller values amplifying differences between candidate scores and larger values making these scores less separated, $w$ and $\tau_t$ are hyperparameters fixed at 0.6 and 30 days, respectively, and the time-compatibility penalty is bounded in implementation for numerical stability. In addition, $t_{o,j}$ is the time stamp of first detection for optical sample $j$, while $t_{g,i}$ is the merger time for GW event $i$. The second term in Equation~\ref{eq:time_bias} encourages matches whose optical reference time is compatible with the queried \gw\ trigger.

Because one \gw\ event can correspond to multiple optical realizations, the supervised contrastive objective \citep{khoslaSupervised2021} treats every optical realization sharing the anchor \gw\ event as a positive. For a \gw\ anchor $i$ with positive set $P(i)$, the objective function is given by
\begin{equation}
\mathcal{L}_{g\rightarrow o,i}
=-\frac{1}{|P(i)|}\sum_{p\in P(i)}
\log
\frac{\exp(s_{ip})}
{\sum_j \exp(s_{ij})},
\label{eq:g2o_loss}
\end{equation}
and we average the loss over both GW$\rightarrow$optical and optical$\rightarrow$GW directions. This branch provides coarse retrieval and stabilizes the subsequent fusion stage.

\begin{deluxetable}{lccccccc}
\tablecaption{Model dimensions used in the main experiments \label{tab:param_value}}
\tablehead{
\colhead{Quantity} & \colhead{$D$} & \colhead{$J$} & \colhead{$H$} & \colhead{$K$} & \colhead{$L_o$} & \colhead{$L_g$} & \colhead{$d$}
}
\startdata
Value & 128 & 128 & 192 & 64 & 64 & 300 & 128 \\
\enddata
\tablecomments{$D$ is the hidden dimension of the GW features. $J$ and $K$ are the dimensions of the optical global vector and coordinate embedding, respectively, and $H$ is the hidden dimension of the optical temporal tokens. $L_o$ and $L_g$ are the lengths of the optical temporal-token sequence and GW localization-token sequence, respectively. The symbol $d$ denotes the key-vector dimension used in the dual cross-attention module.}
\end{deluxetable}

\subsection{Dual cross-attention fusion}

Pure contrastive similarity is insufficient for retrieval in large candidate pools, because it cannot explicitly reason about query-specific spatial and physical consistency. We therefore introduce fine-grained cross-modal fusion with a dual cross-attention mechanism \citep{vaswaniAttention2017,liAlign2021}:
\begin{align}
\bm{f}_{\rm param\rightarrow opt} &=\operatorname{Attn}(\bm{g}_{\rm param}, \bm{H}_{\rm opt}, \bm{H}_{\rm opt}), \\
\bm{f}_{\rm coord\rightarrow gw} &=\operatorname{Attn}(\bm{c}_{\rm opt}, \bm{H}_{\rm gw}, \bm{H}_{\rm gw}),
\end{align}
where the attention computation is given by
\begin{equation}
\begin{aligned}
\operatorname{Attn}&(\bm X_q,\bm X_k,\bm X_v)=\\
&\operatorname{softmax}\!\left(\frac{(\bm X_q\bm W^Q)(\bm X_k\bm W^K)^\intercal}{\sqrt d}\right)(\bm X_v\bm W^V).
\end{aligned}
\end{equation}
In the attention formula, \(\bm X_q\), \(\bm X_k\), and \(\bm X_v\) represent the query, key, and value input matrices, respectively;
\(\bm W^Q\), \(\bm W^K\), and \(\bm W^V\) are the corresponding learnable weight matrices;
and \(d\) is the dimension of each key vector, used for scaling the dot products.

The first branch identifies which aspects of the optical temporal evolution are most compatible with the \gw\ source parameters, while the second branch identifies which regions of the \gw\ localization map are most consistent with the candidate coordinates. The layer-normalized attention outputs are weighted and concatenated with the scaled time separation, $\delta t=\min(|t_o-t_g|/(30\mathrm{d}),10)$, and passed to a classifier. The output logit $l$ serves as the ranking score, and $\sigma(l)$ gives the pairwise match probability.

The two directions of cross-attention correspond to physically distinct queries. Optical coordinates query sky-localization structure, while \gw\ source parameters query light-curve evolution. Before fusion, we assign weights of 1.0 and 2.0 to the spatial-consistency and source-parameter-consistency branches, respectively. The spatial weight is fixed, and the source-parameter-consistency weight is included in the hyperparameter search described in Section~\ref{sec:hpo}.

\subsection{Training strategy}
\label{sec:training_strategy}

The multimodal model is trained with a four-stage training regime. In all stages, one epoch is defined as 100 optimization steps, with each step using one mini-batch of 1024 GW--optical pairs. During the first stage, covering epochs 1--8, the GW and optical encoders and the contrastive temperature are optimized with the supervised contrastive objective $\mathcal{L}_{\mathrm{goc}}$, averaged over both GW$\rightarrow$optical and optical$\rightarrow$GW directions \citep{khoslaSupervised2021}; the fusion head is frozen. This initial alignment stage provides coarse cross-modal retrieval and stabilizes the representations used by the subsequent fusion stages.

During the second stage, covering epochs 9--12, the encoders are frozen and the pairwise classifier is optimized while $\lambda_{\mathrm{cls}}$ is linearly increased to 1.0. We construct four types of sample pairs for the two-class cross-entropy objective $\mathcal{L}_{\mathrm{cls}}$. Matched positive pairs associate a positive \gw\ event with one of its retained \kn\ light curves. Optical-negative pairs associate a positive \gw\ event with a non-\kn\ transient. \gw-negative pairs associate a source-type-matched optical candidate with an ejecta-free or optically undetected \gw\ event. Mismatched pairs associate a positive \gw\ event with a \kn\ light curve generated for a different merger, preferentially from the same source class. 
%We average the cross-entropy loss within each pair category and combine the aligned-positive, negative-GW, mismatched-KN, and non-KN terms with weights 0.50, 0.20, 0.15, and 0.15, respectively, using label smoothing of 0.1.

During the third stage, covering epochs 13--16, the encoders remain frozen, the classification-loss weight is held at $\lambda_{\mathrm{cls}}=1.0$, and the retrieval-loss weight is linearly increased to $\lambda_{\mathrm{ret}}=1.0$. For each selected positive \gw\ query $i$, we construct a mixed training gallery $\mathcal{G}_i$ containing 1000 optical candidates: one associated \kn\ light curve, 250 \kn\ distractors generated for other \gw\ events, and 749 non-\kn\ transients sampled from ELAsTiCC2.

The parent events of the \kn\ distractors are sampled uniformly. A parent may contribute several distinct optical realizations, and each optical row appears at most once within a query gallery. All candidate coordinates are set to the coordinate of the associated counterpart. Each \kn\ candidate retains its first-detection delay relative to its parent \gw\ event. For non-\kn\ candidates, the time delay is sampled with equal probability from the empirical \kn-delay distribution and a uniform distribution over $0$--$30$ days. The same non-\kn\ delay distribution is used for optical-negative classification pairs. Gallery loss is evaluated for at most 128 queries from each mini-batch.

Let $\mathcal{V}$ denote the selected positive-query set. Assuming that the fusion logit for candidate $j\in\mathcal{G}_i$ is $l_{ij}$ and that $p_i$ denotes the associated counterpart, the mixed-gallery objective is
\begin{equation}
\mathcal{L}_{\mathrm{ret}}
=
-\frac{1}{|\mathcal{V}|}
\sum_{i\in\mathcal{V}}
\log
\frac{\exp(l_{i,p_i})}
{\sum_{j\in\mathcal{G}_i}\exp(l_{ij})},
\label{eq:ret_loss}
\end{equation}
This objective trains the fusion logits to concentrate probability mass on the associated counterpart within each mixed gallery.

During the fourth stage, covering epochs 17--100, the encoders and fusion head are jointly fine-tuned. The encoder learning rate is set to 0.1 times the fusion-head learning rate, while $\lambda_{\mathrm{cls}}=1.0$, $\lambda_{\mathrm{ret}}=1.0$, and the effective contrastive weight decreases linearly from 0.5 to 0.25.

The total loss across the four-stage training strategy is
\begin{equation}
\mathcal{L}=
\lambda_{\mathrm{goc}}\mathcal{L}_{\mathrm{goc}}+\lambda_{\mathrm{cls}}\mathcal{L}_{\mathrm{cls}}+\lambda_{\mathrm{ret}}\mathcal{L}_{\mathrm{ret}},
\label{eq:total_loss}
\end{equation}
where $\lambda_{\mathrm{goc}}$, $\lambda_{\mathrm{cls}}$, and $\lambda_{\mathrm{ret}}$ are adjusted according to the four training stages described above.

The model is trained for 100 epochs with a batch size of 1024. Training uses AdamW \citep{loshchilovDecoupled2018}, gradient clipping, weight decay, and a five-epoch learning-rate warmup followed by cosine decay. Validation is performed with gallery sizes of 100, 500, 1000, 2000, and 5000, using 64 BNS queries, 64 NSBH queries, and two gallery trials. The checkpoint score combines source-macro MRR and Recall@1 defined in Eq.~(\ref{eq:recall}) and Eq.~(\ref{eq:mrr}) with weights 0.8 and 0.2 and uses gallery-size weights of 0.10, 0.15, 0.20, 0.25, and 0.30.

\subsection{Hyperparameter optimization}
\label{sec:hpo}

Hyperparameters are optimized exclusively on the training--validation data described in Section~\ref{ch:dataset_split}. Within each source class, the event-level validation set is partitioned into disjoint tuning and confirmation subsets containing 75\% and 25\% of the validation GW events. The tuning subset supports checkpoint selection, trial scoring, and resource allocation. The confirmation subset is reserved for the final multi-seed assessment of leading configurations. The optimization objective is evaluated on mixed validation galleries containing one matched counterpart and KN and non-KN distractors. KN distractors retain their parent-relative first-detection delays and share the counterpart sky coordinates. Non-KN distractors are assigned positions within the 90\% credible localization region and delays drawn equally from the empirical KN-delay distribution and a uniform distribution over 0--30 days.

We use Optuna \citep{akibaOptuna2019} with a multivariate grouped tree-structured Parzen estimator \citep{bergstraAlgorithms2011} to evaluate 50 configurations, including 12 startup trials. To allocate computational resources efficiently, a multi-stage successive-halving schedule \citep{jamiesonNonstochastic2016} trains all 50 configurations for 30 epochs, advances the top 25 configurations to 60 epochs, and evaluates the top 12 configurations to 100 epochs. The six leading feasible configurations are then evaluated with three different random seeds.

The optimization score combines source-macro mean reciprocal rank and Recall@1 across five gallery sizes:
\begin{equation}
\mathcal{J}_{\rm HPO}
=
\sum_{N\in\mathcal{S}} \alpha_N
\left(
0.8\,\overline{\mathrm{MRR}}_N
+0.2\,\overline{\mathrm{R@1}}_N
\right),
\label{eq:hpo_objective}
\end{equation}
where $\mathcal{S}=\{100,500,1000,2000,5000\}$, the overbar denotes an equal-weight average over the BNS and NSBH validation samples, and the gallery-size weights are $\alpha_N=\{0.10,0.15,0.20,0.25,0.30\}$ in ascending order of $N$. At each resource-allocation stage, a candidate model is retained only if its recall is at least 0.85 for each negative-GW category and its validation area under the precision–recall curve (AUPRC) is no more than 0.005 below that of the reference configuration evaluated at the same stage. The reference configuration, using the fiducial values of the search domains, is evaluated under the same training budget at each stage. These conditions prevent gains in retrieval performance from weakening the rejection of GW events without detectable counterparts or degrading overall pairwise classification performance. The final configuration is selected from candidates that provide consistent retrieval gains across seeds while preserving large-gallery MRR, AUPRC, and negative-GW recall. Table~\ref{tab:hpo_search} lists the complete search space, fiducial values and selected values.

\begin{deluxetable*}{lccc}
\tabletypesize{\small}
\tablewidth{0pt}
\tablecaption{Hyperparameter search domains and selected values \label{tab:hpo_search}}
\tablehead{
\colhead{Hyperparameter} & \colhead{Search domain} & \colhead{Selected value} & \colhead{Fiducial value}
}
\startdata
Learning rate & $[3\times10^{-4},\,7\times10^{-4}]$ & $5.78\times10^{-4}$ & $5.00\times10^{-4}$ \\
Weight decay & $[0.01,\,0.03]$ & 0.0214 & 0.0200 \\
Retrieval weight, $\lambda_{\rm ret}$ & $\{0.75,\,1.00,\,1.25\}$ & 1.00 & 1.00 \\
Contrastive weight, $\lambda_{\rm goc}$ & $\{0.50,\,1.00,\,1.50\}$ & 1.00 & 1.00 \\
Classification weight, $\lambda_{\rm cls}$ & $\{0.50,\,0.75,\,1.00\}$ & 1.00 & 1.00 \\
Physical-fusion weight & $\{1.5,\,2.0,\,2.5\}$ & 2.0 & 2.0 \\
Alignment-stage duration & $\{6,\,8,\,10\}$ epochs & 8 & 8 \\
Initial joint-stage multiplier of $\lambda_{\rm goc}$ & $\{0.1,\,0.3,\,0.5\}$ & 0.5 & 0.5 \\
KN distractor fraction & $\{0.10,\,0.25,\,0.40\}$ & 0.25 & 0.25 \\
Training gallery size & $\{500,\,750,\,1000\}$ & 1000 & 1000 \\
\enddata
\tablecomments{The learning rate and weight decay are sampled log-uniformly, and the remaining parameters are sampled categorically. The spatial-consistency fusion weight is fixed at 1.0.}
\end{deluxetable*}

\section{Feasibility Study and Results}

To assess the robustness and effectiveness of MAGIKS, we conduct four sets of experiments on the test dataset.
\begin{enumerate}
    \item \textbf{Retrieval ablation study}: This experiment analyzes the impact of retrieval-oriented training objectives and cross-modal architectures, including fusion methods, loss components, and a reproduced Fink random forest baseline, on candidate ranking capability.
    \item \textbf{GW170817 retrieval experiment}: This experiment evaluates the MAGIKS model for a real event using the released GW source parameters and localization information, the measured source position and host distance, and counterfactual Rubin observations.
    \item \textbf{Classification ablation study}: This experiment evaluates separation of associated GW--KN pairs from optical-negative, GW-negative, and mismatched pairs and examines the contributions of the classification loss and fusion components.
    \item \textbf{GW-parameter sensitivity study}: This experiment uses matched crossed pairs to measure the sensitivity of the MAGIKS score to individual GW parameters while controlling the remaining physical and observational properties. Results are compared with a physics-informed empirical bridge and a GW-independent optical control.
\end{enumerate}

\subsection{Experimental setup}

All ablation experiments use the independent test set described in Section~\ref{ch:dataset_split}. Retrieval queries are drawn from the 1,843 positive GW events, and both negative-GW subclasses contribute to the pairwise-classification evaluation. In retrieval mode, each gallery contains one associated counterpart and $N-1$ distractors. Candidate first detections must occur within 30 days after the GW trigger, and their coordinates must lie inside the 90\% credible region of the GW skymap \citep{petrovDatadriven2022,andreoniTargetofopportunity2022,shahPredictions2024,stevensonStrategy2026}. We evaluate gallery sizes of 10, 100, 500, 1000, 2000, and 5000.

Existing optical follow-up observations of GW events \citep{liuIlluminating2026,gillandersPanSTARRS2025,huKilonova2025,anandLIGO2025} indicate that, after basic filtering of artifacts, stars, and other non-transient sources, the surface density of plausible astrophysical transients is typically of order one candidate per square degree. A localization region of approximately $1000~{\rm deg}^{2}$ can therefore contain approximately $10^{3}$ candidates. The evaluated gallery sizes cover both compact candidate sets and large localization regions. For each GW event, each of ten trials samples one associated light-curve realization and a distractor sequence. Gallery sizes use nested prefixes of that sequence, and the same galleries are shared by all methods. Classification metrics are computed from 40,960 sampled pairs, with 10,240 examples in each of the four pair categories. Events are sampled without replacement within each batch.

We use Recall@K and mean reciprocal rank (MRR) to evaluate retrieval. Assuming there are $M$ queries and the rank of the associated counterpart is $r_i$, the metrics are
\begin{align}
\mathrm{Recall@K} &= \frac{1}{M}\sum_{i=1}^{M}\mathbf{I}(r_i \le K),\label{eq:recall} \\
\mathrm{MRR} &= \frac{1}{M}\sum_{i=1}^{M}\frac{1}{r_i},\label{eq:mrr}
\end{align}
where $\mathbf{I}$ is the indicator function. We further aggregate results across $A$ gallery sizes using a logarithmically weighted average:
\begin{align}
\overline{\mathrm{MRR}} &=
\frac{\sum_{i=1}^{A}\log_{10}(s_i)\,\mathrm{MRR}_i}
{\sum_{i=1}^{A}\log_{10}(s_i)}, \\
\overline{\mathrm{Recall@K}} &=
\frac{\sum_{i=1}^{A}\log_{10}(s_i)\,\mathrm{Recall@K}_i}
{\sum_{i=1}^{A}\log_{10}(s_i)},
\end{align}
where $s_i$ denotes the $i$th gallery size.

The classification experiments contain four pair types. Matched positive pairs associate a positive GW event with one of its associated KN light curves. Optical-negative pairs combine a positive GW event with a non-KN transient. GW-negative pairs combine an optical candidate with an ejecta-free or optically undetected merger. Mismatched pairs combine a positive GW event with a KN light curve generated for another merger. The threshold-dependent metrics are
\begin{align}
\mathrm{Accuracy} &= \frac{TP+TN}{TP+TN+FP+FN}, \\
\mathrm{Precision} &= \frac{TP}{TP+FP}, \\
\mathrm{Recall} &= \frac{TP}{TP+FN}, \\
F_1 &= \frac{2\,\mathrm{Precision}\,\mathrm{Recall}}
{\mathrm{Precision}+\mathrm{Recall}},
\end{align}
where $TP$, $FP$, $TN$, and $FN$ denote the numbers of true positives, false positives, true negatives, and false negatives. The $F_1$ score provides a single measure of the trade-off between precision and recall. We also report two threshold-independent metrics: the area under the receiver operating characteristic curve (AUROC) and the area under the precision--recall curve (AUPRC). The ROC curve is obtained by varying the classification threshold and plotting the true positive rate (TPR, or Recall) against the false positive rate (FPR), where $\mathrm{TPR}=TP/(TP+FN)$ and $\mathrm{FPR}=FP/(FP+TN)$.

\subsection{Retrieval ablations}

We performed the retrieval ablation experiment with multiple model settings to assess the modules and training strategies related to candidate ranking. We compared eight methods: MAGIKS; variants without retrieval loss, dual cross-attention, the fusion branch, or contrastive loss; skymap-only and optical-only baselines; and a reproduced Fink random forest baseline. The latter adapts the Fink kilonova science module \citep{biswasEnabling2023} to noisy LSST $ugrizY$ light curves, using shared principal-component light-curve features and a random-forest classifier. It uses the same training population as MAGIKS and ranks gallery candidates by the predicted kilonova probability without using the \gw\ query. A light curve is scoreable when it contains at least two usable bands, with at least two valid photometric measurements in each band. Under this criterion, 84.7\% of the counterpart light curves and 67.7\% of the contaminant light curves are scoreable. For samples below this threshold, the available photometric information is insufficient for a reliable assessment; no kilonova probability is assigned, and these candidates are placed after scoreable candidates in the gallery ranking. In the variant without cross-attention, the cross-attention fusion mechanism is replaced by concatenating the encoder feature vectors before the classification head. In the variant without the fusion branch, the fusion branch is not trained, and the contrastive similarity is used for ranking. For the skymap-only baseline, we use the credible level of the optical sky coordinates in the \gw\ skymap as the ranking statistic.

The random-forest baseline produces discretized probability scores and therefore exact ties. For this baseline, we report the analytical expectation under uniform random ordering within each tied-score group. If $g$ distractors score strictly above the counterpart and $e$ distractors share its score, the counterpart is assigned equal probability to ranks $g+1,\ldots,g+e+1$; Recall@K and MRR are averaged over these possible ranks. This avoids dependence on input order or on a particular random permutation of tied candidates.

\begin{deluxetable*}{lcccccccccccc}
\tablewidth{\textwidth}
\tabletypesize{\small}
\tablecaption{Summary of retrieval ablation experiments at representative gallery sizes \label{tab:ret_ablation}}
\tablehead{
\colhead{\raisebox{-0.65\normalbaselineskip}{Method}} &
\multicolumn{3}{c}{Gallery = 100} &
\multicolumn{3}{c}{Gallery = 500} &
\multicolumn{3}{c}{Gallery = 1000} &
\multicolumn{3}{c}{Gallery = 5000} \\
\cline{2-4}
\cline{5-7}
\cline{8-10}
\cline{11-13}
\colhead{} &
\colhead{R@1} & \colhead{R@10} & \colhead{MRR} &
\colhead{R@1} & \colhead{R@10} & \colhead{MRR} &
\colhead{R@1} & \colhead{R@10} & \colhead{MRR} &
\colhead{R@1} & \colhead{R@10} & \colhead{MRR}
}
\startdata
Skymap-only & 0.051 & 0.319 & 0.140 & 0.012 & 0.101 & 0.046 & 0.007 & 0.056 & 0.028 & 0.002 & 0.013 & 0.008 \\
Optical-only & 0.577 & 0.870 & 0.689 & 0.321 & 0.753 & 0.475 & 0.211 & 0.677 & 0.367 & 0.048 & 0.396 & 0.149 \\
Fink Random Forest & 0.639 & 0.787 & 0.705 & 0.392 & 0.758 & 0.539 & 0.244 & 0.732 & 0.424 & 0.038 & 0.560 & 0.156 \\
w/o Retrieval loss & 0.510 & 0.883 & 0.637 & 0.324 & 0.676 & 0.438 & 0.260 & 0.571 & 0.362 & 0.144 & 0.356 & 0.218 \\
w/o Cross-attention & 0.279 & 0.554 & 0.378 & 0.153 & 0.391 & 0.234 & 0.110 & 0.320 & 0.180 & 0.039 & 0.180 & 0.084 \\
w/o Fusion & 0.133 & 0.416 & 0.228 & 0.055 & 0.207 & 0.109 & 0.035 & 0.149 & 0.076 & 0.010 & 0.064 & 0.029 \\
w/o Contrastive loss & \underline{0.773} & \underline{0.983} & \underline{0.850} & \underline{0.595} & \textbf{0.891} & \underline{0.700} & \underline{0.515} & \textbf{0.835} & \underline{0.626} & \underline{0.349} & \underline{0.660} & \underline{0.451} \\
MAGIKS & \textbf{0.776} & \textbf{0.985} & \textbf{0.852} & \textbf{0.614} & \underline{0.891} & \textbf{0.710} & \textbf{0.543} & \underline{0.834} & \textbf{0.642} & \textbf{0.398} & \textbf{0.666} & \textbf{0.488} \\
\enddata
\tablecomments{Each gallery contains one true kilonova counterpart and time- and sky-compatible distractors. R@1 and R@10 denote the fractions of queries for which the true counterpart is ranked first or within the top ten candidates, respectively, and MRR is its mean reciprocal rank. For the Fink baseline, metrics are analytical expectations under uniform random ordering within exact-score ties, and candidates with insufficient photometric coverage are placed after scoreable candidates. Best values are shown in bold, and second-best values are underlined. Best and second-best values are determined before rounding;}
\end{deluxetable*}

\begin{figure*}
    \centering
    \includegraphics[width=\textwidth]{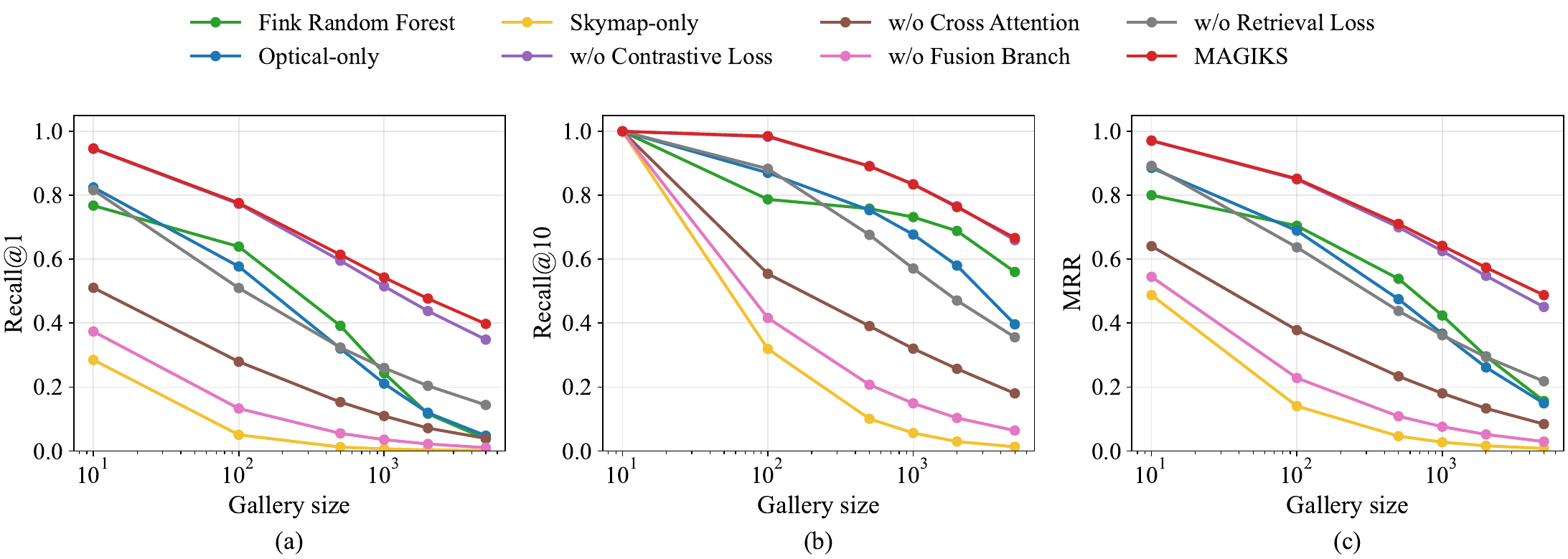}
    \caption{Retrieval performance as a function of gallery size for the single-modality baselines, the reproduced Fink random forest, multimodal ablations, and MAGIKS. Each gallery contains one true kilonova counterpart and time- and sky-compatible distractor transients. Panels (a), (b), and (c) show Recall@1, Recall@10, and mean reciprocal rank (MRR), respectively. Recall@K measures the fraction of queries for which the true counterpart is ranked within the top $K$ candidates, while MRR measures the average inverse rank of the true counterpart. Metrics are averaged over ten independently sampled galleries for each query and gallery size.}
    \label{fig:retrieval_curves}
\end{figure*}

The core retrieval ablation study at four representative gallery sizes is summarized in Table~\ref{tab:ret_ablation} and shown in Figure~\ref{fig:retrieval_curves}. MAGIKS provides the highest R@1 and MRR at all four gallery sizes in Table~\ref{tab:ret_ablation}. At gallery size 1000, MAGIKS reaches Recall@1 = 0.543, Recall@10 = 0.834, and MRR = 0.642, compared with 0.211, 0.677, and 0.367 for the optical-only baseline. At gallery size 5000, MAGIKS reaches Recall@1 = 0.398, Recall@10 = 0.666, and MRR = 0.488, compared with 0.048, 0.396, and 0.149 for the optical-only baseline. The reproduced Fink random forest retains Recall@10 = 0.560 at gallery size 5000, while its Recall@1 and MRR decrease to 0.038 and 0.156. These results show that \gw\ conditioning contributes most clearly to top-rank ordering as the candidate pool grows.

The ablations separate the roles of the model components. Removing the fusion branch causes the largest degradation among the multimodal variants, confirming that the contrastive similarity alone does not provide sufficient query-specific evidence for large-gallery reranking. Removing cross-attention substantially reduces retrieval performance, supporting the effectiveness of the structured dual cross-modal matching mechanism. MAGIKS gives higher Recall@1 and MRR than the variant without contrastive loss at all four gallery sizes. Their Recall@10 values are nearly identical at gallery sizes 500 and 1000, while MAGIKS is higher at 100 and 5000. Removing retrieval loss reduces Recall@10 and MRR at gallery size 5000 from 0.666 and 0.488 to 0.356 and 0.218, respectively. Thus, the fusion architecture and retrieval supervision provide the principal gains in ranking performance, with the contrastive objective further improving top-rank ordering.

\begin{figure*}
    \centering
    \includegraphics[width=\textwidth]{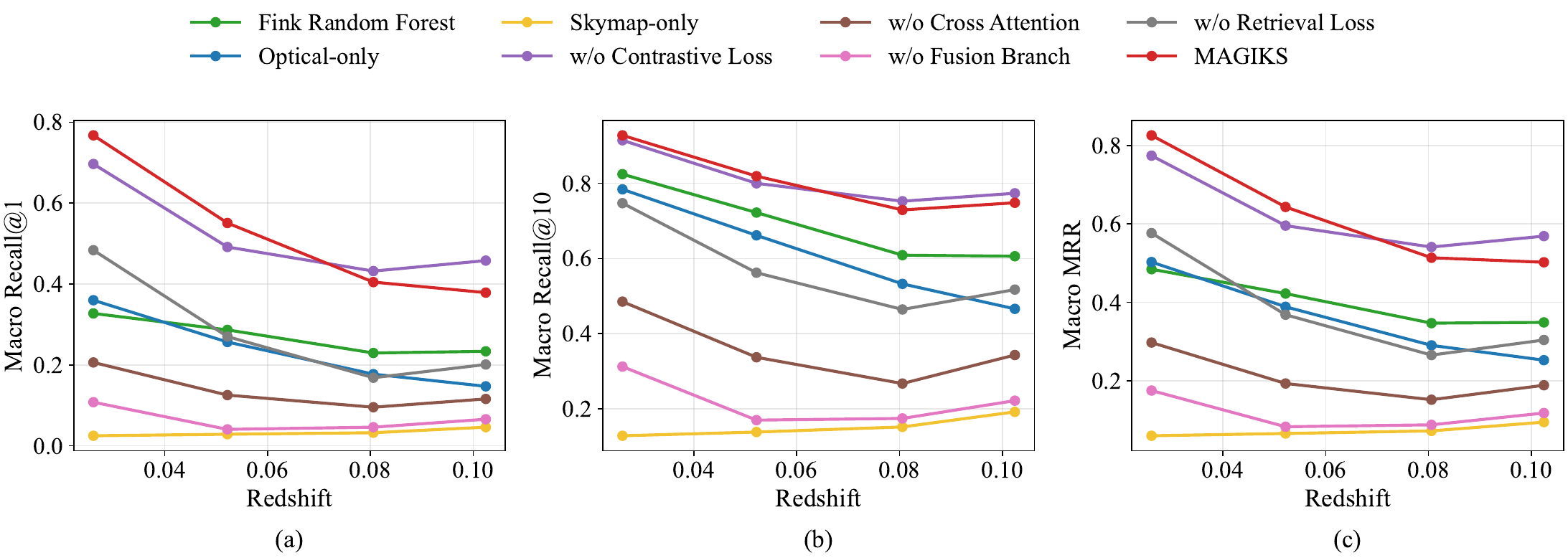}
    \caption{Redshift dependence of retrieval performance for the single-modality baselines, the reproduced Fink random forest, multimodal ablations, and MAGIKS. Positive-query GW events populate four redshift bins, $[0.00,0.04)$, $[0.04,0.065)$, $[0.065,0.10)$, and $[0.10,0.14)$; the plotted values are the mean redshifts of the contributing queries, approximately 0.026, 0.052, 0.081, and 0.102. Panels (a), (b), and (c) show logarithmically weighted macro Recall@1, macro Recall@10, and macro MRR across the six evaluated gallery sizes.}
    \label{fig:retrieval_redshift_curves}
\end{figure*}

We also computed macro metrics at different redshifts to evaluate generalization and performance stability across distance scales, as shown in Figure~\ref{fig:retrieval_redshift_curves}. The selected positive queries populate four redshift bins: $[0.00,0.04)$, $[0.04,0.065)$, $[0.065,0.10)$, and $[0.10,0.14)$, with representative mean redshifts of 0.026, 0.052, 0.081, and 0.102. MAGIKS achieves macro Recall@1 values of 0.768, 0.551, 0.405, and 0.379 across the four populated bins. MAGIKS gives the highest macro Recall@1, Recall@10, and MRR in the first two bins. The variant without contrastive loss gives the highest values for all three metrics in the two higher-redshift bins. The reproduced Fink random forest provides macro Recall@10 values of 0.824, 0.722, 0.609, and 0.606, exceeding the optical-only baseline in this metric but remaining below the two strongest multimodal settings. Removing cross-attention or the fusion branch produces substantial reductions across all populated bins. These results show that the multimodal fusion mechanism remains effective across the evaluated redshift range, while the contribution of the contrastive objective is not uniform across distance scales.


\subsection{GW170817 retrieval under Rubin observing scenarios}
\label{sec:gw170817_retrieval}

We evaluate a GW170817 query using the released event information and counterfactual Rubin observations. The detector-frame component masses are set to the medians of the GWTC-1 low-spin posterior, $m_{1}^{\rm det}=1.479\,M_{\odot}$ and $m_{2}^{\rm det}=1.281\,M_{\odot}$, with aligned spin components $\chi_{1z}=0.00291$ and $\chi_{2z}=0.00144$ and $\cos\iota=-0.835$ \citep{ligoscientificcollaborationandvirgocollaborationGWTC12019,ligoscientificcollaborationandvirgocollaborationProperties2019}. The GW input retains the released three-dimensional BAYESTAR sky and distance posterior, whose distance mean and standard deviation are 34.15 and 9.34 Mpc, respectively \citep{abbottMultimessenger2017,richabbottOpen2021}. The optical simulations include the measured AT2017gfo position, R.A. $=197.450374^{\circ}$ and decl. $=-23.381495^{\circ}$, and use the NGC 4993 host distance of 40.7 Mpc \citep{cantielloPrecise2018}.

Ten observing epochs are generated by annual sidereal shifts of the GW170817 trigger within the Rubin baseline. Each epoch uses a shared panel of 50 coordinates that includes the measured source position and locations sampled inside the 90\% credible region. The resulting $10\times50$ panel separates variation due to survey epoch from variation across the localization region. Each scenario--coordinate light curve is evaluated in three independent galleries at sizes 10, 100, 500, 1000, 2000, and 5000. We average the three repeats and estimate 95\% confidence intervals from 10,000 two-way bootstrap samples, independently resampling observing epochs and coordinates with replacement \citep{efronBootstrap1979}.

\begin{deluxetable*}{lccc}
\tablewidth{0pt}
\tabletypesize{\small}
\tablecaption{MAGIKS retrieval for the GW170817 experiment \label{tab:gw170817_ret}}
\tablehead{
\colhead{Gallery size} & \colhead{R@1 [95\% CI]} & \colhead{R@10 [95\% CI]} & \colhead{MRR [95\% CI]}
}
\startdata
10   & $0.990\,[0.964,1.000]$ & $1.000\,[1.000,1.000]$ & $0.994\,[0.980,1.000]$ \\
100  & $0.964\,[0.891,0.999]$ & $0.995\,[0.980,1.000]$ & $0.975\,[0.921,1.000]$ \\
500  & $0.941\,[0.863,0.991]$ & $0.977\,[0.920,1.000]$ & $0.955\,[0.882,0.995]$ \\
1000 & $0.931\,[0.850,0.983]$ & $0.970\,[0.902,1.000]$ & $0.947\,[0.870,0.991]$ \\
2000 & $0.905\,[0.820,0.967]$ & $0.961\,[0.889,1.000]$ & $0.930\,[0.851,0.980]$ \\
5000 & $0.873\,[0.781,0.947]$ & $0.957\,[0.880,1.000]$ & $0.906\,[0.826,0.966]$ \\
\enddata
\tablecomments{Point estimates are averages over ten Rubin observing epochs, 50 coordinates, and three gallery repeats per scenario--coordinate cell. Confidence intervals are percentiles from 10,000 two-way bootstrap replicates over the observing-epoch and coordinate axes.}
\end{deluxetable*}

MAGIKS preserves strong top-rank retrieval as the gallery grows. At a gallery size of 1000, the model reaches R@1 of 0.931, R@10 of 0.970, and MRR of 0.947. At 5000 candidates, R@1 is 0.873, R@10 is 0.957, and MRR is 0.906. The two-way intervals quantify the finite variation across the ten observing epochs and the shared coordinate panel.

\subsection{Classification ablation experiments}
\label{sec:classification_ablation}

To examine the contribution of the classification components, we compare four models: MAGIKS and variants without dual cross-attention, without the fusion branch, or without classification loss. In the last variant, $\lambda_{\rm cls}$ remains zero throughout training. The comparison uses the matched-positive, optical-negative, GW-negative, and mismatched pair definitions described above.

\begin{figure*}
    \centering
    \includegraphics[width=0.68\textwidth]{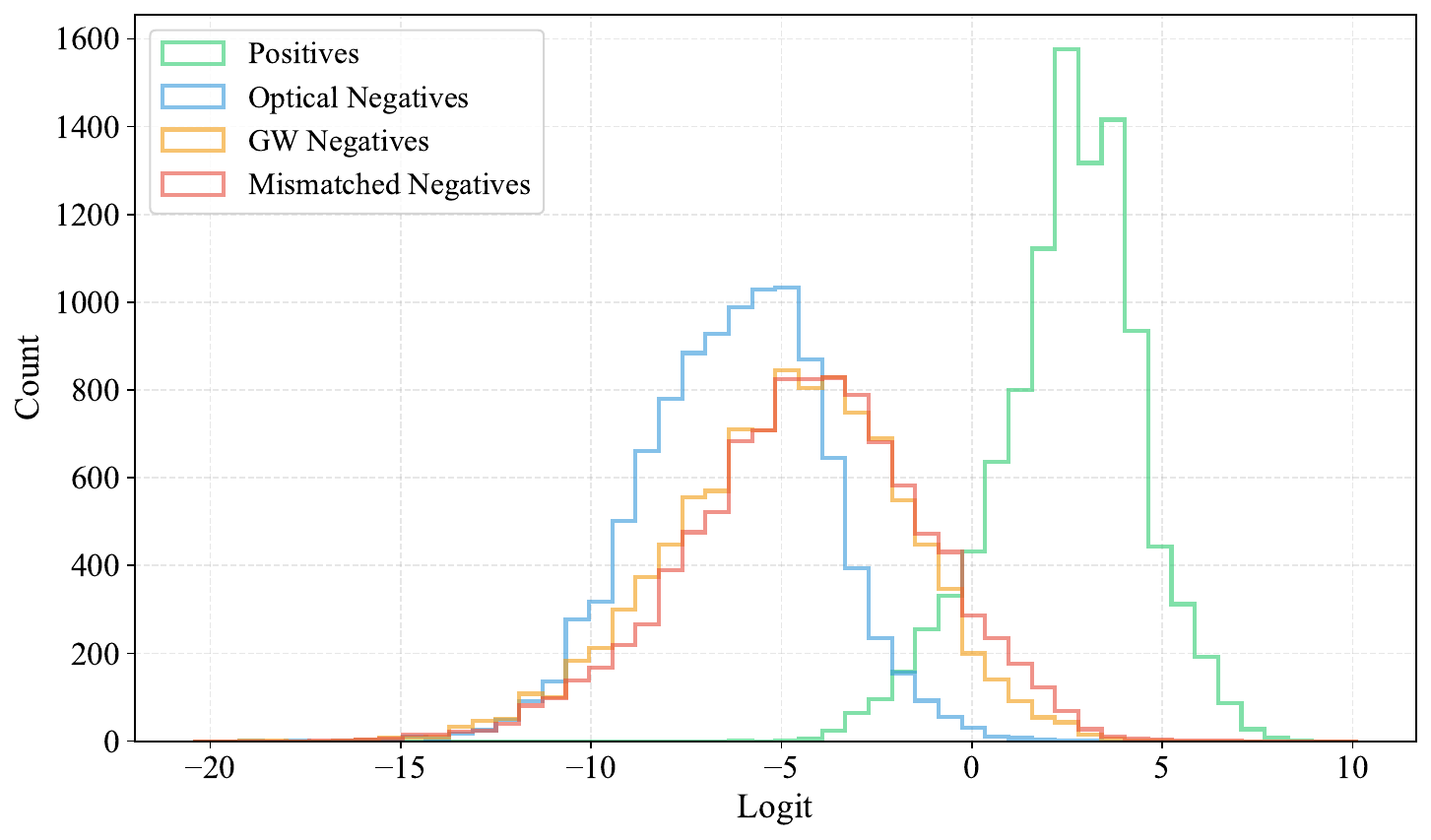}
    \caption{Classification-logit distributions for matched positive pairs and three negative-pair categories. Optical-negative pairs combine a positive GW event with a non-KN transient, GW-negative pairs combine an optical candidate with an ejecta-free or optically undetected merger, and mismatched pairs combine a positive GW event with a KN light curve generated for another merger. Larger logits indicate stronger support for an associated pair.}
    \label{fig:logits_distribution}
\end{figure*}

Figure~\ref{fig:logits_distribution} shows the classification-logit distributions for the four pair categories. The logit is converted to a positive-match probability through $\operatorname{sigmoid}(x)=(1+\exp(-x))^{-1}$. Matched positive pairs are concentrated at high logits, while most negative pairs receive lower scores. The high-logit tails of the GW-negative and mismatched categories constitute the most challenging negative-pair cases.

\begin{deluxetable}{lcccccc}
\tablewidth{\linewidth}
\tabletypesize{\normalsize}
\tablecaption{Summary of pairwise-classification ablation experiments \label{tab:cls_ablation}}
\tablehead{
\colhead{Method} & \colhead{AUROC} & \colhead{AUPRC} & \colhead{Accuracy} & \colhead{Precision} & \colhead{Recall} & \colhead{$F_1$}
}
\startdata
w/o cross-attention & \underline{0.978} & \underline{0.938} & \underline{0.928} & 0.810 & \textbf{0.929} & \underline{0.865} \\
w/o fusion & 0.555 & 0.294 & 0.700 & 0.328 & 0.190 & 0.241 \\
w/o classification loss & 0.929 & 0.818 & 0.864 & \underline{0.834} & 0.567 & 0.675 \\
MAGIKS & \textbf{0.983} & \textbf{0.951} & \textbf{0.942} & \textbf{0.877} & \underline{0.893} & \textbf{0.885} \\
\enddata
\tablecomments{The positive class contains associated GW--KN pairs. The negative class combines optical-negative, GW-negative, and mismatched pairs. AUROC and AUPRC are threshold-independent measures; accuracy, precision, recall, and $F_1$ use the common fixed threshold of 0.5. Best values are shown in bold, and second-best values are underlined.}
\end{deluxetable}

Table~\ref{tab:cls_ablation} shows that MAGIKS provides the strongest overall classification performance, with AUROC of 0.983, AUPRC of 0.951, accuracy of 0.942, precision of 0.877, and $F_1$ of 0.885. The variant without cross-attention reaches the highest recall, 0.929, but has lower AUROC, AUPRC, accuracy, precision, and $F_1$ than MAGIKS.

Without classification loss, AUROC decreases to 0.929 and recall to 0.567, producing an $F_1$ score of 0.675. Removing the complete fusion branch causes the largest degradation, with AUROC of 0.555 and $F_1$ of 0.241. These results show that query-specific cross-modal fusion is central to pairwise separation and that the classification objective improves the balance between precision and recall at the adopted threshold.

\begin{figure*}
    \centering
    \includegraphics[width=0.82\textwidth]{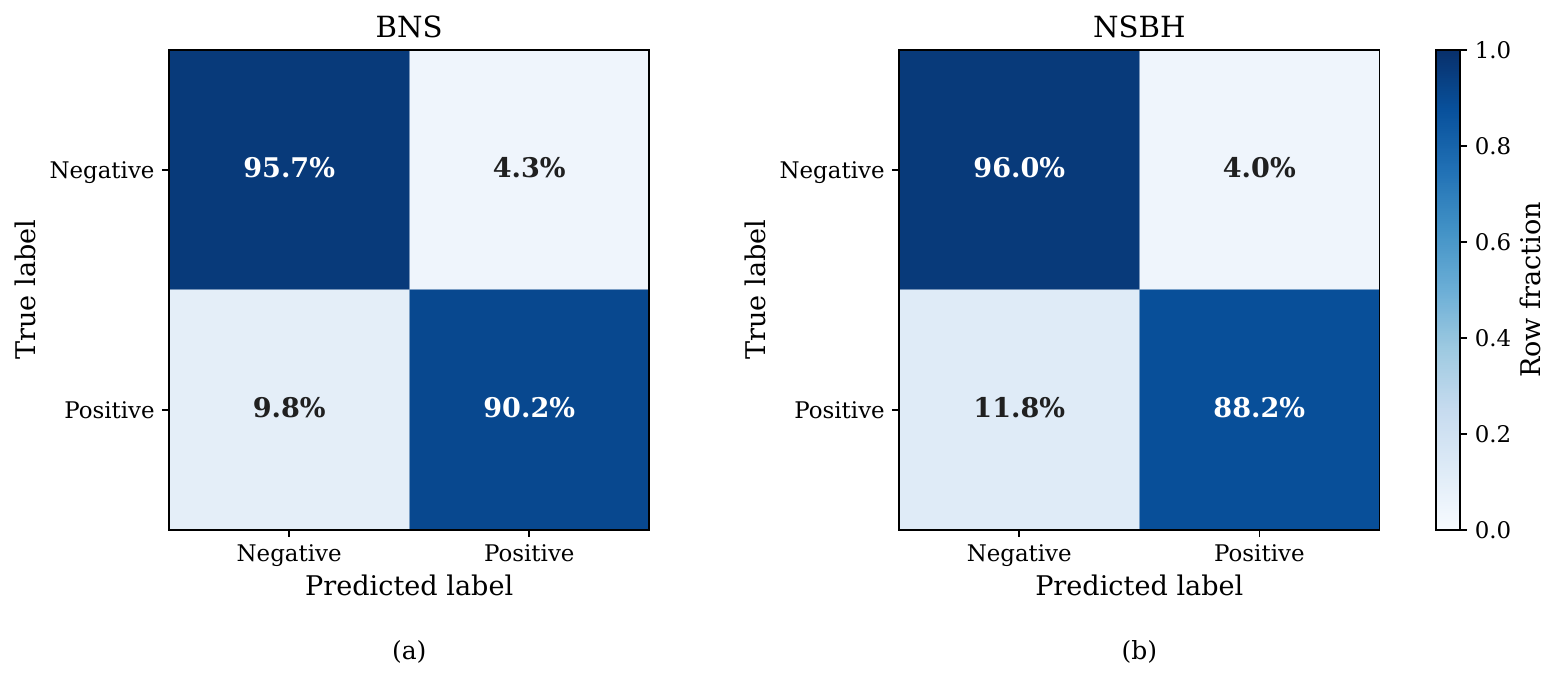}
    \caption{Row-normalized confusion matrices for MAGIKS pairwise classification of BNS and NSBH systems. The positive class contains associated GW--KN pairs, and the negative class combines optical-negative, GW-negative, and mismatched pairs.}
    \label{fig:CM_BNS_NSBH}
\end{figure*}

MAGIKS also retains high true-negative and true-positive rates for both source populations, as shown in Figure~\ref{fig:CM_BNS_NSBH}. For BNS pairs, the false-positive and true-positive rates are 4.3\% and 90.2\%, respectively. The corresponding NSBH rates are 4.0\% and 88.2\%.

\subsection{Sensitivity to individual GW parameters}
\label{sec:gw_parameter_sensitivity}

We use matched crossed pairs of events to test whether the matching score responds consistently to differences in an individual GW parameter while approximately controlling the remaining parameters and optical sampling. The targets are detector-frame chirp mass $\mathcal{M}_{c}^{\rm det}$, mass ratio $q$, inclination $|\cos\iota|$, luminosity distance $\log_{10}(d_L/{\rm Gpc})$, and the NSBH primary aligned-spin component. The adopted BNS ejecta prescription is spin-independent, so spin sensitivity is evaluated only for NSBH systems. For NSBH systems, the more massive primary is the black hole, so we use its aligned spin $\chi_{\rm BH,z}=\chi_{1z}$. This is the spin that controls the ISCO radius and therefore directly affects the adopted NSBH ejecta prescription.

For each target and source class, we require the paired events to differ by at least one interquartile range (IQR) in that target. Every other coordinate in the source-specific matching basis must differ by no more than 0.5 IQR, while the fractional distance uncertainty and optical sampling summaries are also caliper matched. We then pair light-curve realizations according to their number of observations, band coverage, and temporal span over $-10$ to $+20$ days relative to merger. Within each crossed comparison, both light curves receive the same sky coordinate and trigger-relative time offset. We repeat the comparison using each event as the common anchor and average over anchors and matched light-curve realizations before averaging independent event pairs.

For two matched events $A$ and $B$, we evaluate a $2\times2$ table of scores: $s_{AA}=\operatorname{Score}(\mathrm{GW}_A,\mathrm{Optical}_A)$ and $s_{BB}$ are the physically associated pairs, whereas $s_{AB}$ and $s_{BA}$ exchange their optical counterparts. For query $A$, the score is directionally correct when $s_{AA}>s_{AB}$; for query $B$, it is correct when $s_{BB}>s_{BA}$. We average these two wins over the $P$ independent event pairs,
\begin{equation}
R_{\rm dir}=\frac{1}{2P}\sum_{p=1}^{P}
\left[h(s_{AA}-s_{AB})+h(s_{BB}-s_{BA})\right],
\end{equation}
where $h(x)$ equals 1 for $x>0$, 0.5 for $x=0$, and 0 for $x<0$. The crossed-pair design controls for preferences based on either modality alone. A GW-only offset cancels within each query, while a query-independent optical preference favours opposite directions in the two comparisons. Thus, an additive score $s(\mathrm{GW},\mathrm{Optical})=u(\mathrm{GW})+v(\mathrm{Optical})$ gives $R_{\rm dir}=0.5$, making the experiment sensitive to cross-modal dependence.

We construct a physics-informed empirical bridge as a data-driven reference for the physical associations accessible to MAGIKS. Without using a deep-learning model, the bridge maps the GW query and the optical light curve independently into a common physical state,
\begin{equation}
\boldsymbol{\psi}=\left(
\log_{10}\frac{M_{\rm dyn}}{M_\odot}),\,
\log_{10}\frac{M_{\rm wind}}{M_\odot}),\,
|\cos\iota|,\,
\log_{10}\frac{d_L}{\mathrm{Gpc}}
\right).
\end{equation}
This state combines the ejecta masses governing the intrinsic emission with the viewing orientation and luminosity distance that determine its observed appearance \citep{coughlinMultimessenger2019,foucartRemnant2018,krugerEstimates2020,raaijmakersChallenges2021}.

On the GW side, kernel-weighted nearest neighbours in the training set estimate the ejecta-state distribution from $(\mathcal{M}_{c}^{\rm det},q)$ for BNS systems and $(\mathcal{M}_{c}^{\rm det},q,\chi_{\rm BH,z})$ for NSBH systems. The query supplies the inclination and distance, with the distance uncertainty propagated into the state covariance. On the optical side, a separate neighbour search estimates the same physical state from light-curve summaries of brightness, temporal evolution, colour, and cadence. Each side is represented by a Gaussian distribution with a neighbour-weighted mean and covariance. The matching score is the negative Bhattacharyya distance between these distributions, so stronger agreement in the inferred physical state produces a higher score \citep{loftsgaardenNonparametric1965,kailathDivergence1967}.

\begin{figure*}
    \centering
    \includegraphics[width=0.86\textwidth]{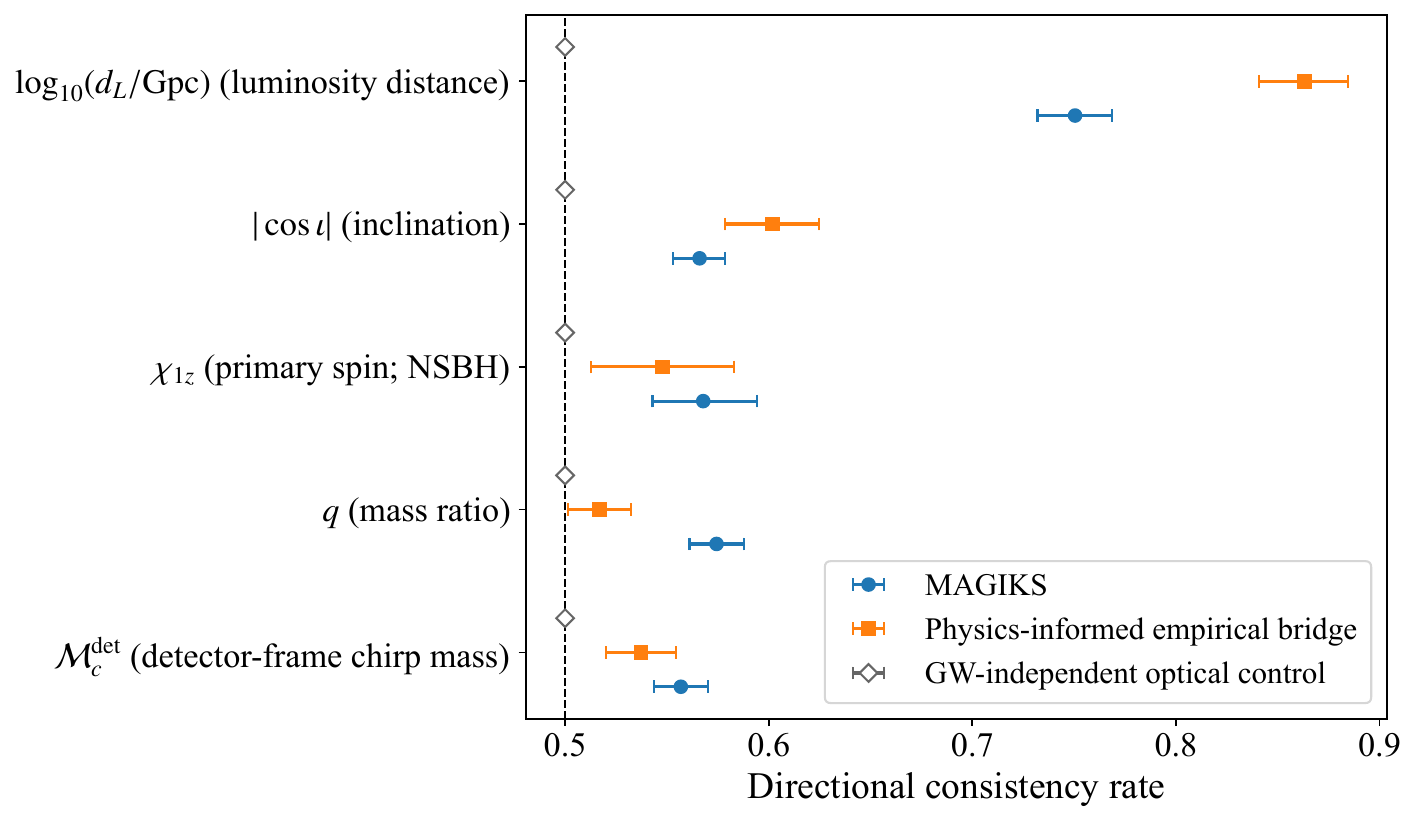}
    \caption{Directional consistency rates for MAGIKS, the physics-informed empirical bridge, and the GW-independent optical control in the single-parameter crossed-pair experiment. Points show the primary 0.5-interquartile-range caliper estimates, and error bars give 95\% intervals from 10,000 bootstrap samples. The vertical dashed line marks the GW-independent value of 0.5. Detector-frame chirp mass, mass ratio, inclination, and luminosity distance are averaged with equal weight over BNS and NSBH sources. Primary black-hole spin is evaluated for NSBH systems, for which the ejecta prescription is spin dependent.}
    \label{fig:gw_parameter_sensitivity}
\end{figure*}

MAGIKS gives directional consistency rates above the GW-independent control for all five reported parameters, shown in Figure~\ref{fig:gw_parameter_sensitivity}. It captures substantial associations with luminosity distance and inclination, with rates of 0.750 and 0.566, although these remain below the bridge estimates of 0.863 and 0.602. For NSBH primary spin and detector-frame chirp mass, MAGIKS reaches 0.568 and 0.557, close to the bridge values of 0.548 and 0.537. Mass ratio shows a larger excess over the bridge, 0.574 versus 0.517. This may reflect limited recovery of mass ratio in the matched-filter estimates \citep{ligoscientificcollaborationandvirgocollaborationProperties2019}, weakening its reliable physical association with the optical signal; explicitly conditioning MAGIKS on this uncertain parameter may introduce spurious correlations.

\subsection{Discussion}
\label{sec:discussion}

The experiments highlight several complementary strengths of \model. First, multimodal retrieval provides clear advantages over single-modality ranking, especially when the gallery size becomes large. The optical-only baseline and reproduced Fink random forest method remain competitive in small and moderately sized galleries because the light-curve morphology carries strong counterpart information, but their performance degrades faster than MAGIKS as more time- and sky-compatible distractors are added. The skymap-only baseline is useful as a coarse spatial prior, but it provides limited discriminative power for fine ranking. By conditioning optical candidates on \gw\ localization and source-parameter information, \model\ maintains substantially higher MRR and Recall@10 in the large-gallery regime, showing that the \gw\ modality contributes most strongly when counterpart prioritization becomes operationally difficult.

Second, the ablation results separate the roles of the major architectural and training components. Removing the fusion branch causes the largest degradation, indicating that shared contrastive embeddings alone are insufficient for query-specific ranking. Replacing cross-attention with feature concatenation also substantially weakens retrieval performance. MAGIKS gives higher Recall@1 and MRR at all four gallery sizes than the variant without contrastive loss, while Recall@10 is nearly identical between the two models. The no-retrieval-loss variant shows that pairwise classification supervision can generate useful ranking scores, but its lower large-gallery metrics indicate that explicit retrieval supervision improves top-rank ordering.

The auxiliary classification experiments provide complementary evidence for this interpretation. Although retrieval is the primary task, pairwise classification probes whether the model learns a separable matching function between \gw\ signals and optical candidates. \model\ achieves the best AUROC, AUPRC, accuracy, and $F_1$ score among the classification ablations, and maintains low false-positive rates for both \bns\ and \nsbh\ source types. The variant without classification loss still shows partial separability, but its lower recall and $F_1$ score indicate that the classification objective improves calibration and fixed-threshold discrimination. Thus, retrieval and classification supervision play complementary roles: retrieval loss structures the ranking space, while classification loss improves pairwise decision reliability.

The GW170817 experiment provides an event-specific assessment using the released GW source parameters and three-dimensional localization, together with the measured counterpart position and host distance. Across counterfactual Rubin observing conditions, MAGIKS maintains effective candidate prioritization as the gallery becomes increasingly crowded. This stability indicates that the learned GW--optical matching function remains informative under changes in survey epoch and position within the localization region. However, the experiment is based on a single event with simulated Rubin light curves and therefore does not capture uncertainty from alternative GW posterior samples, kilonova emission models, or real alert-stream contaminants.

The single-parameter sensitivity study provides complementary evidence that the MAGIKS score uses information from the GW query. Its directional consistency exceeds the GW-independent optical control for all tested parameters. The bridge comparison indicates substantial learned distance and inclination dependence and comparable sensitivity for chirp mass and NSBH spin. These results demonstrate that MAGIKS captures meaningful GW--optical associations across multiple physical parameters, with the enhanced mass-ratio sensitivity motivating further investigation of the learned correlations.

In real GW observations, the inclination posterior can be broad, whereas our simulated inputs use the injected inclination. We therefore tested how inclination errors affect retrieval performance. In a supplementary evaluation using a separate model checkpoint and 1,239 test events, we generated 20 independent Gaussian perturbations of the injected inclination angle for each GW event, with 90\% of the angular errors lying within $\pm20^\circ$. All other inputs and model settings were held fixed. At a gallery size of 1000, averaging the retrieval metrics over the 20 realizations reduced Recall@1 from 38.698\% to 38.661\%, a decrease of 0.036 percentage points. Recall@10 decreased by 0.027 percentage points, and MRR decreased by 0.00038. Although this controlled test does not reproduce the full inclination posterior of an observed GW event, the small changes in retrieval performance demonstrate robustness to inclination errors of the tested magnitude under the evaluated conditions.

Relative to prior \gw\ counterpart prioritization approaches based primarily on sky maps, galaxy catalogs, transient filtering heuristics, or single-modality classifiers \citep{singerRapid2016,ducoinOptimizing2020,stachieUsing2020,biswasEnabling2023,liangKilonovaTargeting2023}, \model\ learns an explicit GW--optical matching function and applies pairwise reranking. This is important for future wide-field surveys, where the main challenge is not only detecting transient candidates but also ordering many plausible candidates under limited follow-up resources. The framework can be extended to other classes of optical counterparts by training on their characteristic light-curve populations.

All the experiments were conducted on a single NVIDIA A100 GPU. The model can be trained in less than one day, while the complete workflow, including data generation, preprocessing, training, and evaluation, can be completed in approximately one week. This computational efficiency facilitates its adaptation to other wide-field surveys, such as the Wide Field Survey Telescope (WFST) \citep{wangScience2023,liuTargetofOpportunity2023}.  Excluding data preprocessing and loading, the model achieved a median latency of 5.32 ms per sample pair with one CPU core. At a batch size of 1024, it processed approximately 34,997 samples s\(^{-1}\), with a median batch latency of 29.26 ms on GPU and peak allocated GPU memory of 5.19 GB.

\section{Conclusions}

We present MAGIKS for GW-triggered kilonova counterpart search. The framework combines an encoder for irregular multiband optical time series, a dual-stream GW encoder, time-aware supervised contrastive alignment, mixed-gallery retrieval training, and dual cross-attention fusion. Retrieval and classification ablations evaluate the contribution of different modules and training methods, while the GW170817 and single-parameter experiments provide an event-specific application and tests of physically motivated cross-modal sensitivity.

The results support the following three conclusions within this controlled simulation setting. First, optical-only ranking provides a useful but incomplete reference for counterpart search. Second, dual cross-attention fusion provides query-specific evidence beyond single-modality and contrastive-only scores. Third, the combination of dual cross-attention and retrieval-oriented supervision improves retrieval in large simulated candidate galleries, including gallery sizes up to 5000. Taken together, these findings support the use of \gw-conditioned reranking as a promising direction for future counterpart-prioritization pipelines, provided that robustness to low-latency localization uncertainty, real-time signal identification, and less controlled candidate galleries is tested in subsequent work.

Future work should extend this framework toward more realistic multimessenger follow-up systems. One direction is to broaden the current benchmark by evaluating additional transient-ranking and counterpart-prioritization methods under a common gallery-retrieval protocol. A second direction is external validation with real \gw\ events and real transient candidate streams, where broker pre-filtering can be applied to remove artifacts, Solar System objects, and obvious non-transients before multimodal reranking \citep{mollerFinkNewGeneration2021}. A third direction is to propagate posterior samples for component masses, aligned spins, inclination, and luminosity distance through the \gw\ encoder, enabling uncertainty-aware multimodal matching. The present simulations use the HLV network; future inclusion of KAGRA in an HLVK configuration could improve sky localization and source-parameter recovery when KAGRA contributes appreciable signal-to-noise ratio. The same retrieval formulation could also be generalized beyond Rubin by incorporating survey-specific cadence, depth, and filter information from other wide-field facilities, such as WFST, enabling future wide-area or all-sky kilonova searches with coordinated optical survey networks. Finally, host-galaxy information, galaxy-catalog constraints, and additional messengers such as gamma-ray, X-ray, and radio observations can be integrated into a single multimodal framework to move from candidate prioritization to richer physical interpretation of counterpart populations \citep{parkerAION12025,yingNeural2025}.

\section*{Code Availability}

The source code for \model, including the data processing,
training, and evaluation pipelines, is publicly available at
\url{https://github.com/Super-nova2/gw-kn-multimodal}.

\begin{acknowledgments}
This work is supported by the Fundamental and Interdisciplinary
Disciplines Breakthrough Plan of Ministry of Education
of China, the Strategic Priority Research Program of the Chinese Academy of Science under Grant No. XDB0550300 and the National Natural Science Foundation of China under Grant No. 12325301. This research made use of the OzSTAR national supercomputing facility hosted by Swinburne University of Technology, which is supported by the Australian Government through the National Collaborative Research Infrastructure Strategy (NCRIS) and by the Victorian Government through the Victorian Higher Education State Investment Fund (VHESIF). This work also relied on computational resources provided by the LIGO Laboratory, supported by the U.S. National Science Foundation as a major facility under Grant Nos. PHY-0757058 and PHY-0823459, and was additionally supported by the JST ASPIRE Program of Japan under Grant No. JPMJAP2320.

We thank Sunil Choudhary for helpful guidance on LVK conventions, and Anais Möller and Emille Ishida for valuable advice on reproducing the Fink random forest method. 
\end{acknowledgments}

\begin{contribution}
%%This section gives authors the space to recognize author contributions. The text inside this environment is NOT counted towards the total word quanta. At a minimum, manuscripts are expected to include this text:

Bingzhou Gao and Weichangfeng Guo led the data generation, model development, experiments, and manuscript writing. Jingqi Zhang conducted a subset of the experiments. Linqing Wen and Wen Zhao supervised the project and contributed to the scientific interpretation and manuscript revision. Alistair Mcleod completed the generation of gravitational wave data and provided advice on the manuscript. All authors contributed to the discussion of results and manuscript revision.

%% But authors are expected to provide more specific details, e.g.
%%
%%SC was responsible for writing and submitting the manuscript.
%%WWM came up with the initial research concept and edited the manuscript.
%%OTS obtained the funding and edited the manuscript.
%%EBF provided the formal analysis and validation. He also edited the manuscript.
%%GEH Supervised the undergraduates, wrote the software and administers the project github and Zenodo repositories.
%%
%% Authors can use the Contributor Role Taxonomy (CRediT) at
%% https://credit.niso.org
%% for ideas on how write a good statement tailored to their needs.

\end{contribution}

%% To help institutions obtain information on the effectiveness of their
%% telescopes the AAS Journals has created a group of keywords for telescope
%% facilities.
%
%% Following the acknowledgments section, use the following syntax and the
%% \facility{} or \facilities{} macros to list the keywords of facilities used
%% in the research for the paper.  Each keyword is check against the master
%% list during copy editing.  Individual instruments can be provided in
%% parentheses, after the keyword, but they are not verified.
% \facilities{Rubin}

%% Similar to \facility{}, there is the optional \software command to allow
%% authors a place to specify which programs were used during the creation of
%% the manuscript. Authors should list each code and include either a
%% citation or url to the code inside ()s when available.
\software{GWSamplegen \citep{mcleodBinary2025},
Astropy \citep{2013A&A...558A..33A,2018AJ....156..123A,2022ApJ...935..167A},
PyTorch\citep{paszkePyTorch2019},
Jupyter\citep{kluyverJupyter2016},
Numpy\citep{vanderwaltNumPy2011},
Matplotlib\citep{hunterMatplotlib2007},
Pandas\citep{mckinneyData2010},
HEALPix\citep{gorskiHEALPix2005},
ligo.skymap\citep{singerRapid2016,singerGOING2016,singerSUPPLEMENT2016}
}

%% Appendix material should be preceded with a single \appendix command.
%% There should be a \section command for each appendix. Mark appendix
%% subsections with the same markup you use in the main body of the paper.
%%
%% Each Appendix (indicated with \section) will be lettered A, B, C, etc.
%% The equation counter will reset when it encounters the \appendix
%% command and will number appendix equations (A1), (A2), etc. The
%% Figure and Table counter will not reset.

% \appendix

%% For this sample we use BibTeX plus aasjournalv7.bst to generate the
%% the bibliography. The sample7.bib file was populated from ADS. To
%% get the citations to show in the compiled file do the following:
%%
%% pdflatex sample7.tex
%% bibtext sample7
%% pdflatex sample7.tex
%% pdflatex sample7.tex

\bibliography{reference}{}
\bibliographystyle{aasjournalv7}

%% This command is needed to show the entire author+affiliation list when
%% the collaboration and author truncation commands are used.  It has to
%% go at the end of the manuscript.
%\allauthors

%% Include this line if you are using the \added, \replaced, \deleted
%% commands to see a summary list of all changes at the end of the article.
%\listofchanges

\end{document}